\documentstyle[aas2pp4]{article}

\newcommand{\lograt}{log (N$_{\rm Lyc}$/L$_{\rm FUV})$~}
\newcommand{\ha}{H$\alpha$ {}}
\newcommand{\hans}{H$\alpha$}

\newcommand{\ea}{{\em et~al.} {}}
\newcommand{\eans}{{\em et~al.}}
\newcommand{\msol}{M$_{\odot}$}
\newcommand{\flx}{erg (cm$^{2}$\AA~s) $^{-1}$ {}}

\slugcomment{Submitted to \apj, January 1999}

\lefthead{Stewart et al.}
\righthead{Star Formation Triggering Mechanisms}

\begin{document}

\title{Star Formation Triggering Mechanisms in Dwarf Galaxies:\\
    The Far-Ultraviolet, H$\alpha$, and HI Morphology of Holmberg II}

\author{Susan G. Stewart\altaffilmark{1}}
\affil{U. S. Naval Observatory, 3450 Massachusetts Ave NW,  
Washington, DC 20392-5420}

\author{Michael N. Fanelli}
\affil{Dept. of Physics, University of North Texas, Denton, TX 76203}

\author{Gene G. Byrd}
\affil{Dept. of Physics and Astronomy, University of
  Alabama, Tuscaloosa, AL 35487}

\author{Jesse K. Hill}
\affil{Hughes STX, 4400 Forbes Blvd., Lanham, MD 20706}

\author{David J. Westpfahl}
\affil{Dept. of Physics, New Mexico Institute of Mining \& Technology,
   Socorro, NM 87801}

\author{Kwang-Ping Cheng}
\affil{Dept. of Physics, California State University,
Fullerton, CA 92634}

\author{ Robert W. O'Connell}
\affil{University of Virginia, P.O. Box 3818, Charlottesville, VA 22903}

\author{Morton S. Roberts}
\affil{National Radio Astronomy Observatory, Edgemont Rd.,
   Charlottesville, VA 22903}

\author{Susan G. Neff, Andrew M. Smith, and Theodore P. Stecher}
\affil{Laboratory for Astronomy and Solar Physics, NASA/GSFC,
  Greenbelt, MD 20771}

%\and    

% Notice that each of these authors has alternate affiliations, which
% are identified by the \altaffilmark after each name.  The actual alternate
% affiliation information is typeset in footnotes at the bottom of the
% first page, and the text itself is specified in \altaffiltext commands.
% There is a separate \altaffiltext for each alternate affiliation
% indicated above.

\altaffiltext{1}{NASA Space Grant Fellow, University of Alabama 1995-1997}

% The abstract environment prints out the receipt and acceptance dates
% if they are relevant for the journal style.  For the aasms style, they
% will print out as horizontal rules for the editorial staff to type
% on, so long as the author does not include \received and \accepted
% commands.  This should not be done, since \received and \accepted dates
% are not known to the author.

\begin{abstract}
Far-ultraviolet (FUV), H$\alpha$, and HI observations of dwarf galaxy
Holmberg~II are used to investigate the 
means by which star formation propagates in galaxies lacking global 
internal triggering mechanisms such as spiral density waves. 
The observations trace the interaction between sites of massive star formation and 
the neutral and ionized components of the surrounding ISM
in this intrinsically simple system.
Both local and large-scale triggering mechanisms related to massive star formation
are seen, suggesting that feedback from massive stars is a microscopic process 
operating in all galaxies to a certain degree. 

The data emphasize the importance 
of local conditions in regulating star formation from
evidence such as massive stars inside ionized 
shells, compact HII regions surrounding aging clusters, and stars 
formed in chains of progressing age. Surface brightness 
profiles show current activity correlates 
with the time averaged level of past star formation at a given radius
demonstrating a reliance on local conditions. 
Large-scale triggering by HI 
shells is supported by observations of progenitor populations 
as well as secondary sites of star formation 
associated with their dense rims. Analysis of the 
energy available
from massive stars inside HI shells indicates energy deposited 
into the ISM from supernovae and stellar winds is sufficient to 
account for the HI morphology. Ages of individual star forming 
regions are derived using B, H$\alpha$, and FUV photometry and show both older, 
diffuse FUV regions and younger, compact HII regions. 
The distribution of ages is reconciled with the 
HI morphology, showing a clear preference of young regions for
areas of dense HI and old regions for HI voids. 
Global kinematical properties may also play a role in the 
star formation process since differences in the rotation characteristics
of the neutral gas disk correlate with differences in triggering 
mechanisms. Large-scale feedback from massive stars is shown to 
operate in regions that lack differential shear in the gas disk.

\end{abstract}


\section{Introduction}
The pervading questions surrounding our current 
view of large-scale star formation are primarily associated 
with the need to understand the mechanism for triggering  
and globally maintaining star formation.
The first step in forming a fundamental 
theory to describe star formation in all environments is understanding the process in 
intrinsically simple systems. 
This study utilizes a unique observational dataset to explore the processes 
which may influence new star formation and its relation to the ISM within a 
relatively uncomplicated dynamical environment.
Far-ultraviolet (FUV), H$\alpha$, and HI 
images of dwarf galaxy
Holmberg II (HoII) are used to gain new insight into the star formation process  
by studying the interconnection between massive star formation and the ionized and
neutral components of the galaxy's ISM. 

Irregular galaxies lack a dominant internal global 
triggering mechanism for cloud compression and star formation such as density waves,
yet exhibit a wide range of star formation activity.  
Non-interacting low surface brightness late-type galaxies, including
Magellanic-type irregulars (Im), are 
ideal for this study due to the limited number of global internal processes  
influencing star formation while at the same time having star formation
rates per unit area similar to much more massive spiral galaxies  
(Hunter 1997).
Due to their small size, the amount of energy input per a given
star formation event has a pronounced effect on the global ISM. 
Instabilities in the ISM introduced by massive star formation may play an important role in 
the star formation process in these systems. 
They display relatively uncomplicated internal gas dynamics, generally characterized as slow solid 
body rotation. These characteristics create an environment of reduced 
shear allowing features in the ISM to be longer lived. They often possess huge gas envelopes 
extending well beyond the optical regime of the galaxy. Large vertical scale heights allow 
features in the ISM to grow to large sizes before breaking out of the HI layer. In addition, 
these systems allow for the test of star formation models at very low metallicities. For 
these reasons, late-type low surface brightness galaxies are powerful tools
for understanding the star formation process and galaxy evolution. 
  
The apparently non-interacting Im-type galaxy
HoII (DDO 50, UGC 4305) is a member of the M81-NGC 2403 group 
of galaxies. The galaxy is classified as a David Dunlap
Observatory (DDO) dwarf galaxy due to its low surface brightness
(van den Bergh 1959). Although it is slightly too bright to fit the magnitude
criterion of a ``true dwarf'' galaxy of M$_{B}\gtrsim-$16 (Nilson 1973), it is 
usually referred to as a dwarf galaxy in the literature.
IRAS observations of HoII suggest a minimal presence of dust, 
so little hidden star formation is taking place (Hunter \ea 1989). 
The galaxy has a low metal 
abundance, Z/Z$_{\odot}\sim$0.4 (Hunter \& Gallagher 1985), similar
to that of the Large Magellanic Cloud (LMC).
Hodge \ea (1994) find that the sizes of HII regions 
as well as their size distribution are statistically normal in comparison
to galaxies of the same type, distance, and luminosity.   
HoII has a HI envelope extending to 1.5 D$_{\rm Ho}$, where the Holmberg 
diameter (D$_{\rm Ho}$) is the diameter defined by the B isophote at 26.6 
mag arcsec$^{-2}$. The HI morphology of HoII is intriguing due to the 
various shells and holes present throughout the galaxy (Puche \ea 1992).
Hoessel \ea (1998) estimate a 
distance of 3.05 Mpc using 28 Cepheid candidates. This is different
than the distance estimate of 3.63 Mpc determined by Freedman \ea (1994) 
for M81 using Hubble Space Telescope observations of Cepheid variables. 
Since HoII is likely the closest member of the M81-NGC 2403 group, 
3.05 Mpc is adopted for the distance of HoII. 
Table 1 summarizes the parameters assumed 
for the galaxy in this study. 

%
%---------------TABLE 1---------------------------------------
%
\begin{table*}
\begin{center}
\begin{tabular}{lllr}
\tableline\tableline
Parameter & & Value & Reference \\ 
\tableline
Optical Center &$\alpha_{\rm 1950}$ & 8$^{\rm h}$ 13$^{\rm m}$ 53\fs5 & Dressel \& Condon  (1976) \\
               &$\delta_{\rm 1950}$ & 70\arcdeg 52\arcmin 13\arcsec &  \\
Dynamical Center & $\alpha_{\rm 1950}$ & 8$^{\rm h}$ 13$^{\rm m}$ 53$^{\rm s}$ & Westpfahl (1997) \\
                 & $\delta_{\rm 1950}$ & 70\arcdeg 52\arcmin 47\farcs3 & \\
Distance        &   & 3.05 Mpc & Hoessel, Saha, \& Danielson (1998) \\
M$_{\rm B}$     &   & $-$16.52  & RC3, de Vaucouleurs \ea (1991) \\
Inclination     &   & 41\arcdeg & Puche \ea (1992) \\
Position Angle  &   & 177\arcdeg & Puche \ea (1992) \\
Axial Ratio, q=b/a & & 0.74 & Huchtmeier \& Richter (1988) \\
A$_{\rm B}$         &   & 0.10 & RC3, de Vaucouleurs \ea (1991) \\
A$_{\rm FUV}$       &   & 0.20 & This study \\
\tableline
\end{tabular}
\caption{Adopted Parameters for Holmberg II \label{Table 1}}
\end{center}
\tablenum{1}
\end{table*}

Observations in the vacuum ultraviolet (UV, $\lambda\lambda$ 912 - 3200 \AA) 
isolate hot sources from the underlying optical
cool star background in galaxies. Ultraviolet observations allow direct 
detection of young massive stars, which are responsible for 
the majority of photoionization, photodisassociation, elemental synthesis, 
and kinetic energy input in galaxies. The UV 
observations used in this study are in the FUV window 
($\lambda\lambda$912 - 2000 \AA) and are particularly useful in
characterizing high-mass star formation
since the dominant luminosity sources at 1500 \AA\ are OB stars with mass
M $\gtrsim$ 5 \msol (Fanelli \ea 1992). 
Understanding the UV morphology and its relation to the star 
formation process has broad implications considering that the rest frame
FUV is detectable at high redshift, z$\gtrsim$3, with optical and IR
instruments. O'Connell (1997) stresses that determining 
a ``morphological k-correction'' for galaxies is needed to assess the 
morphology of distant objects. Therefore, understanding the processes 
responsible for the 
UV emission in nearby systems will play an important role in advancing
our understanding of galaxy evolution.  

This study addresses several issues which Hunter (1997) stresses are vital 
to an improved understanding of star formation processes within irregular 
galaxies. These key issues are tied to the need for a
better understanding of the regulation mechanism 
for star formation on global and local scales, including a
quantitative description of both the feedback processes operating 
in irregular galaxies and the impact which huge stellar complexes have on 
the surrounding ISM. Morphological comparisons between the FUV flux and other 
radiation linked to star formation provide valuable information regarding 
the interaction between a star forming site and its surrounding medium.  
For example, the ratio of the FUV and \ha emission from a young star 
forming region is indicative of the cluster age. The suggested link
between massive star formation and HI 
shells can be studied via the relative FUV and 
HI emission morphologies. Since the FUV is able to isolate the 
massive stellar component over timescales similar to the kinematic
ages of HI shells, this comparison
could identify a progenitor population. Understanding if characteristic
differences in the ISM are related to differences in star formation
mechanisms is another important issue. The FUV image can be used
to probe the possible relationship between massive star formation and
the kinematics of its HI disk. The intrinsic dynamical reduction 
of shear in the gas disks compared to spiral galaxies most likely 
influences the overall star formation rate. 

Despite the wealth of information provided
by UV images, they are seldomly used in star formation studies because 
spatially-resolved UV images are rare. The FUV images used in this
study are provided by the Ultraviolet Imaging Telescope (UIT) which flew 
as a member of the ASTRO Observatory on two Shuttle missions in 
December, 1990 and March, 1995. Together, a total of $\sim 35$ FUV images of 
galaxies exhibiting recent star formation were observed. 
The $\sim 3\arcsec$ spatial resolution of the UIT   
was a 5$-$20 times improvement over previously 
available UV imagery.  
Besides the Apollo 16 UV image of the LMC in 1972 (Page \& Carruthers 1981), the only other 
UV image of a galaxy similar in type to HoII prior to the ASTRO-2 mission was the dwarf 
galaxy Holmberg IX. It was serendipitously 
observed during the ASTRO-1 mission due to its close proximity to M81.  
Using the UV data, Hill \ea (1993) characterized the most recent star formation event, 
confirming that no recent star formation was occurring in Holmberg IX. 
During the second flight of the UIT, a dwarf galaxy program 
was initiated. The results of this program were first presented by 
Gessner \ea (1995). Detailed analysis of all of the UIT dwarf galaxy targets
is presented by Stewart (1998). 

Global photometry of the galaxy is given in \S3 and the radial surface
brightness profiles in \S4. Photometry of individual star forming complexes
along with their derived ages and internal extinctions are 
presented in \S5. Comparisons between the relative morphologies of the 
FUV and \ha emission are given in \S6 and that of the FUV and HI in \S7.
The principal results are summarized in \S8.

\section{Observations and Data Reduction}

Holmberg II was observed by the UIT during the ASTRO-2 mission 
in March, 1995. Table 2 summerizes the observations of the galaxy used in this study.
The targets were observed using the 
broad-band UIT B1 filter, which has $\lambda_{eff} = 1521$ \AA\ and 
$\Delta\lambda$ = 354 \AA. The images were recorded on IIaO 
film and digitized with a PDS microdensitometer at Goddard
Space Flight Center, producing 2048$\times$2048 pixel 
images with scale 1\farcs14 pixel$^{-1}$. 
The images were calibrated using IUE spectrophotometry of
stars observed by the UIT. The FUV magnitudes
are computed from FUV flux using m$_{FUV}$= $-2.5\times$ log$_{10}$(FUV)$-$21.1,
where FUV is the incident flux in erg (cm$^{2}$\AA~s) $^{-1}$. The chief photometric 
uncertainties are in the form of low-level nonuniformities introduced via the 
development and digitization processes. Uncertainty in the absolute calibration 
can lead to uncertainties of up to 10 $-$ 15~\% in the FUV flux. 
A detailed discussion of the reduction of the UIT 
data to flux-calibrated arrays is given by Stecher \ea (1992, 1997).  

%
%---------------TABLE 2---------------------------------------
%
\begin{table*}
\begin{center}
\begin{tabular}{lccl}
\tableline\tableline
Telescope & Filter & Exposure     & Date\\
          &        & Time (s)     &     \\
\tableline
UIT & FUV (B1) & 1310  & Mar 11, 1995\\
MLO & U        & 1200  & Nov ~2, 1995 \\
MLO & B        & 300   & Feb 24, 1995 \\
MLO & R        & 120   & Feb 24, 1995 \\
MLO & \ha      & 600   & Feb 24, 1995 \\

\tableline
\end{tabular}
\end{center}
\tablenum{2}
\caption{Summary of Observations for Holmberg II \label{Table 2}}
\end{table*}

The FUV background level ($\mu >$ 25 mag arcsec$^{-2}$)
is checked by taking the mean 
of 20, 20$\times$20 pixel boxes in areas void of any galaxy or stellar
flux. The FUV background level is minimal, 0.0$\pm0.4$ analog data units (ADU's). 
The UIT has limited accuracy at the lowest light levels. The error in the background level 
reflects this intrinsic uncertainty. 

CCD observations of HoII were obtained in the Johnson U and B bands, 
and the Kron-Cousins R and \ha bands with San Diego State University's 
1.0 m telescope at the Mount Laguna Observatory (MLO). 
The optical images are median combined with at least two other images
having the same exposure time to remove cosmic rays and other defects. 
Astrometry and photometry are implemented using standard IDL procedures
for data reduction. Background values are determined for each image 
by taking the mean of pixels in 20, 10$\times$10 pixel boxes. 
The HoII broad-band images are  
calibrated using unpublished photoelectric photometry  of three bright 
stars in the galaxy (Corwin 1997).  Total magnitudes given in the RC3 agree with our measured values 
within reasonable limits providing a check of the calibration constants.  

The \ha filter at MLO has a centroid wavelength
6573 \AA\ and width 61~\AA\, which includes the [N~II] lines. 
A correction for this contribution to the \ha flux is made by assuming
f~(\hans)~/~(f~(\ha)~+~f~(~[N~II]~))~$\approx 0.88$, based  on spectral 
observations of H~II regions by Hunter \& Gallagher (1985).
This correction is applied using
the calibration technique of Waller (1990) and observations of the \ha standard
star BD +8\arcdeg2015 (Stone 1977). Attempting to reproduce the irregular
apertures defined by Hodge \ea (1994) 
for H~II regions in HoII provides a check for the \ha  
calibration constant. Our photometry of these regions agree within 
$\sim9$~\%. Deviations may be caused by 
failure to reproduce the irregular apertures exactly.
An \ha emission 
image is  produced by scaling stars in 
the \ha image with those in the R-band image and subtracting the stellar 
continuum component. 

The 21 cm images were obtained with the VLA B, C, and D arrays. The uniformly
weighted column density map has a spatial resolution of 
4$\arcsec\times4\farcs5$, corresponding to a spatial resolution
of 60 pc at the distance of HoII. This image is used in the 
Puche \ea (1992) study of HoII where a complete description of 
its observational characteristics is given. 

\section{Global Photometry}

In Figure 1a the 10\farcm0~$\times$ 10\farcm0 FUV image of HoII
is presented. The FUV morphology is quite patchy with the 
brightest features contained in a central star forming arc. There is 
a region of diffuse FUV emission extending to the southwest and  
three separate patches of FUV emission to the north of the central arc.

Since galaxies at high redshift are seen in the rest frame FUV with optical 
instruments, the FUV image is compared to 
the R-band image in Figure 1b to illustrate the importance of understanding the FUV 
morphology of local galaxies. The R-band image of the galaxy exhibits a relatively 
smooth distribution of light compared to the irregular distribution shown by the 
FUV image. 
The true optical morphology of a high redshift galaxy could be quite different than 
that observed in the optical bands.

\begin{figure}
\plotone{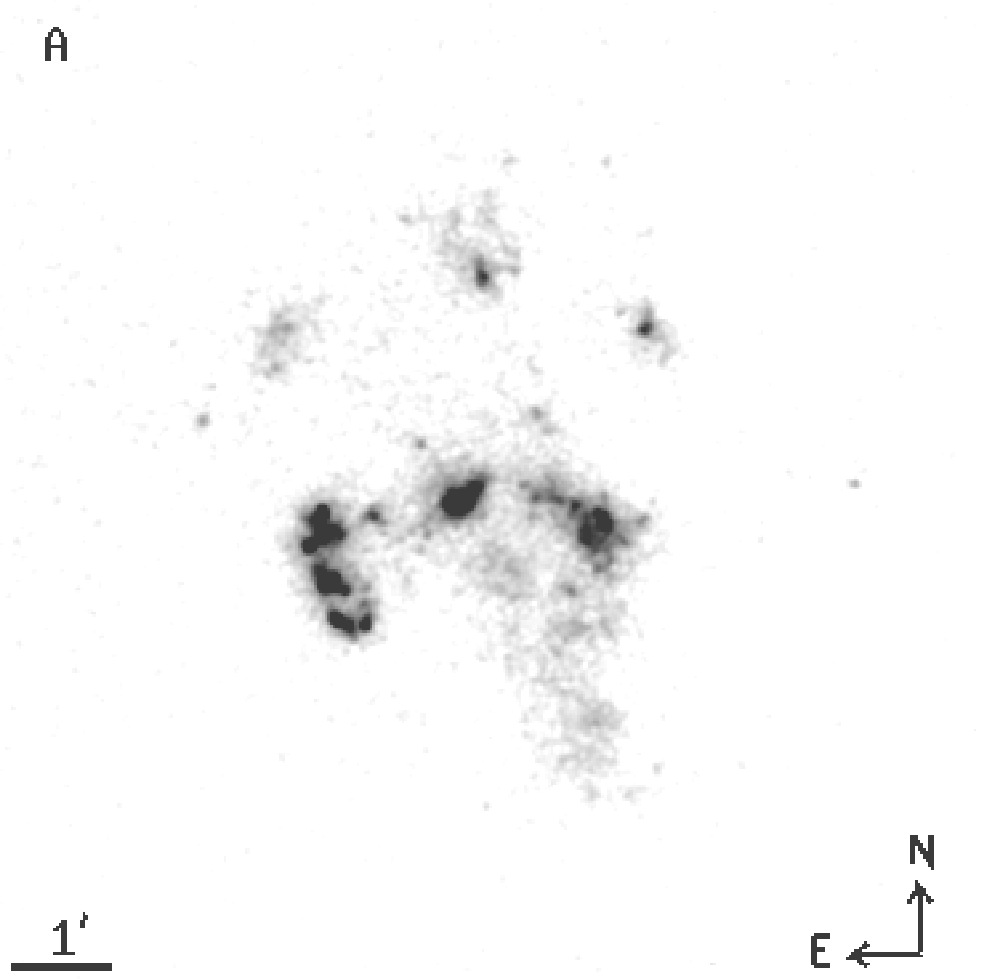}\\
\plotone{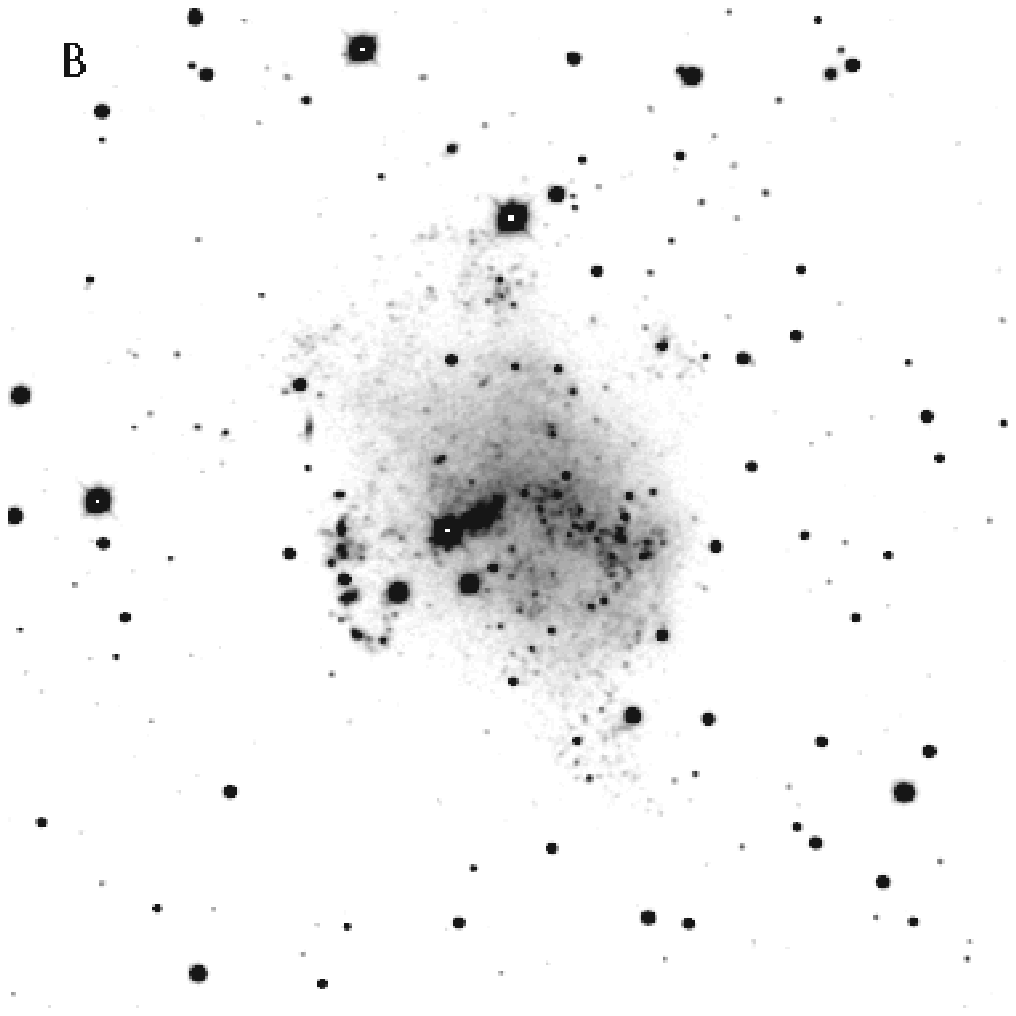}
\caption{Registered FUV (A, top) and R-band images (B, bottom)
of Holmberg~II. \label{Fig 1}}
\end{figure}

Integrated magnitudes of HoII are obtained using elliptical apertures 
whose shape and orientation are based upon the derived HI structural parameters 
given in Table 1. The dynamical center is adopted as the 
aperture center since it has more physical 
significance than the optical center, especially when comparing integrated
parameters from observations ranging from the R-band to the FUV.
The optical magnitudes are derived using images with the 
foreground stars masked, the brightest of which are given 
the average flux from the surrounding medium. Therefore, the  
derived magnitudes may be slightly less than published catalog values. 
Since the FUV image is free from foreground star contamination,
foreground stars are masked from optical images in order to make 
a true comparison with the FUV band. The total integrated magnitudes, corrected 
for Galactic extinction, A$_{\rm g}$, are given for each bandpass in Table 3. 
Galactic extinction in the FUV is derived 
assuming the Galactic reddening curve of Savage and Mathis (1979) integrated over the  
UIT B1 bandpass. These are  related by A$_{\rm FUV}$ / E(B$-$V) = 8.33 (Hill \ea 1997). 
The errors account for the photometric uncertainty and  $\sim$7\% 
uncertainty in the distance estimate. Corrections for internal extinction will
be discussed later (\S5.2).

%
%---------------TABLE 3---------------------------------------
%
\begin{table*}
\begin{center}
\begin{tabular}{lccl}
\tableline\tableline
Filter & Apparent  & Absolute  &A$_{\rm g}$ \\
       & Magnitude & Magnitude &            \\
\tableline
FUV &   9.84$\pm$0.03& $-17.58\pm 0.15$&0.20 \\
  U &  11.48$\pm$0.02& $-15.94\pm 0.15$&0.12 \\
  B &  11.58$\pm$0.02& $-15.84\pm 0.15$&0.10 \\
  R &  10.86$\pm$0.01& $-16.56\pm 0.15$&0.06 \\

\tableline
\end{tabular}
\end{center}
\tablecomments{Magnitudes are corrected for Galactic extinction. 
Optical magnitudes are derived with the foreground stars masked.}
\tablenum{3}
\caption{Integrated Magnitudes for Holmberg II \label{Table 3}}
\end{table*}

The FUV magnitude is at the faint 
end of the range of absolute magnitudes observed by the UIT. In a sample 
of 35 galaxies exhibiting recent massive star formation, the range of
observed absolute magnitudes spans from $-$17 to $-$22  (Fanelli \ea 1997b). 
The corresponding FUV$-$B =~$-$1.74$\pm$0.03 color falls 
into the range observed by FAUST (a balloon-born 
UV camera) for a sample of Im-type galaxies (Deharveng \ea 1994).

The integrated FUV flux as a function of radius  provides information regarding 
the concentration of recent massive star formation.
The FUV growth curve, corrected for Galactic extinction, is shown in 
Figure 2. It illustrates that  
recent massive star formation is not particularly centrally concentrated in HoII
despite the presence of the prominent central star forming arc. 
The aperture containing this feature in its entirety (1\farcm7)   
only contains $\sim$43\% of the total FUV flux. Therefore, massive star formation 
extends well beyond the central star forming arc. 

\begin{figure}
\plotone{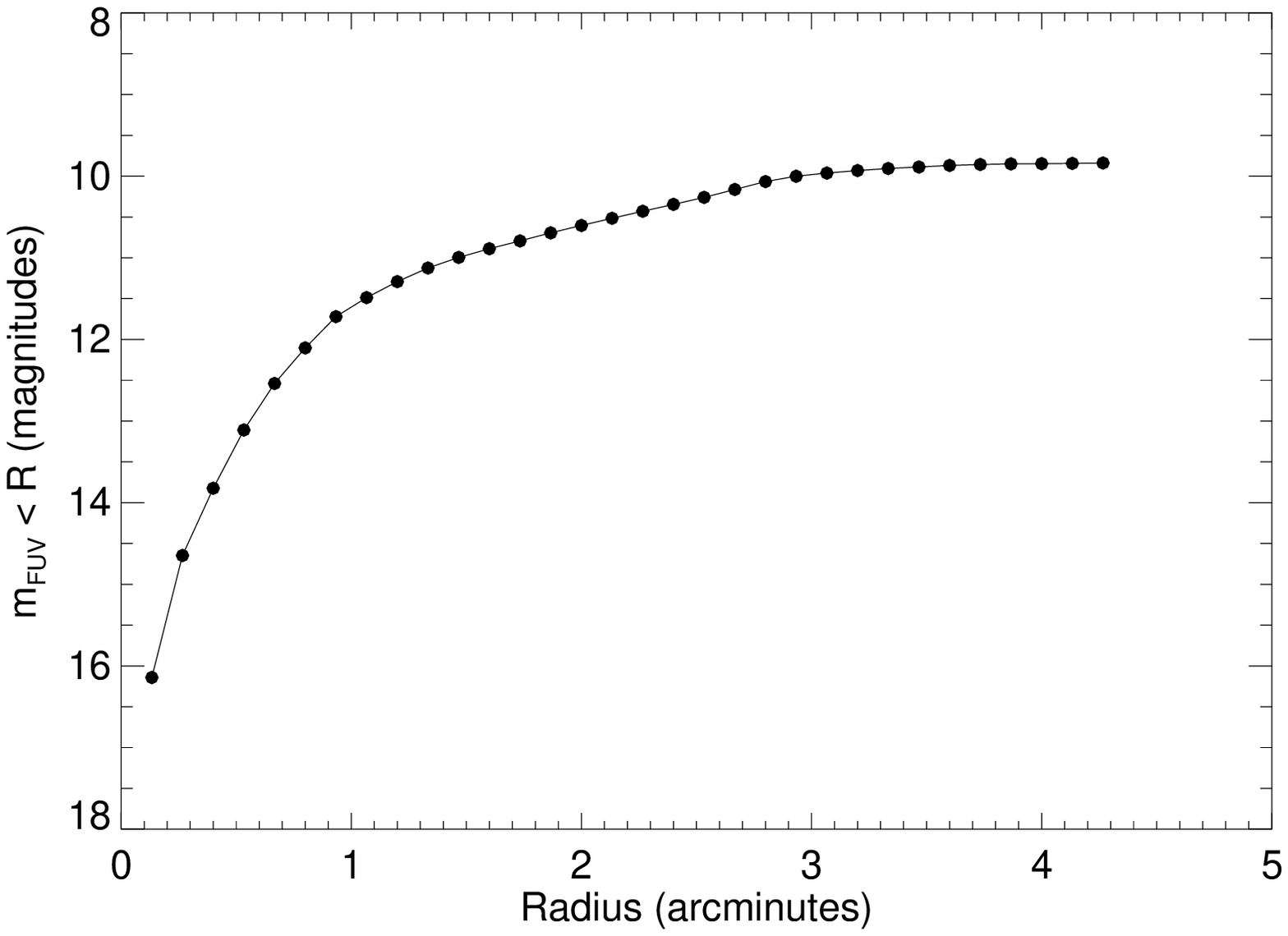}
\caption{The FUV growth curve of Holmberg II. \label{Fig 2}}
\end{figure}

\subsection{Global Star Formation Rates}

The intrinsic emission of a young stellar population 
can be translated into a global star formation rate (SFR) by comparing 
integrated fluxes with predictions from stellar population synthesis models. 
The timescale characterized by a given SFR depends upon the 
bandpass of the observation since the luminosity of a single generation 
of stars decays roughly as a power law (O'Connell 1997). 
The Lyman continuum (Balmer emission lines) describes the SFR over a 
timescale $\lesssim$ 5 Myr while the UV continuum characterizes the 
SFR over a 100 Myr timescale.  
Since massive stars (M$\gtrsim$10\msol) are the ionizing sources for the 
Lyman continuum, observed here via the \ha recombination line, 
and the UV emission is dominated by stars $\gtrsim$ 5\msol, the SFR 
derived using these bandpasses describes the formation history of massive stars. 
Late-type irregular galaxies are dominated by short-lived massive stars  
and are therefore excellent sources to provide an estimate of the SFR 
using either FUV or \ha observations.

The global SFR for massive stars in HoII is derived using both the integrated
FUV and \ha flux by applying formulae which convert the   
integrated light from each bandpass to an estimate of the global SFR. These 
formulae are taken from 
Kennicutt (1998) who provides a self-consistent set of formulae for various regimes
 based on the calibration of Madau \ea (1998) for a 
fixed Salpeter (1955) IMF with mass limits 0.1 to 100 \msol.  
The observed (not corrected for internal extinction) FUV and \ha luminosity and 
global SFR are listed in Table 4. The \ha luminosity and SFR$_{\rm H\alpha}$  
agree reasonably well with the values derived by Miller \& Hodge (1994) after 
differences in distances are removed. 

Since internal extinction is not extreme low metallicity systems, 
the comparison between SFR$_{\rm H\alpha}$ and  SFR$_{\rm FUV}$ is valid to
the first order in estimating the recent to current formation history of massive 
stars. For example, differences in internal extinction effects between the FUV and \ha  
observations for a given E(B-V) are likely on the order of  
A$_{\rm FUV}$/A$_{\rm H\alpha}\sim $4, assuming a LMC extinction  
curve for internal reddening (this ratio doubles for the SMC
extinction curve). This difference translates into 
a higher SFR$_{\rm FUV}$ than SFR$_{\rm H\alpha}$ in a given system by 
underestimating A$_{\rm FUV}$ to a greater degree than A$_{\rm H\alpha}$.   
The results indicate the ``recent'' 
(traced by the FUV) and ``current'' (traced by \ha) massive star formation rates
are comparable.
The ratio SFR$_{\rm FUV}$ to SFR$_{\rm H\alpha}$ indicates
that the FUV flux traces a slightly larger 
fraction of massive star formation than the H$\alpha$ flux, perhaps suggesting 
HoII is in a post-burst phase. 
%
%
%---------------TABLE 4---------------------------------------
%
\begin{table*}
\begin{center}
\begin{tabular}{ccccc}
\tableline
Log(L $_{\rm H \alpha}$) & Log(L$_{\rm FUV}$)&SFR$_{\rm H\alpha}$ & SFR$_{\rm FUV}$  
& SFR$_{\rm FUV}$/SFR$_{\rm H\alpha}$\\ 
                         &   & \msol yr$^{-1}$  & \msol yr$^{-1}$ &        \\  
\tableline\tableline          
  39.55 & 38.67 & 0.028 & 0.049 & 1.75    \\ 
\tableline
\end{tabular}
\end{center}
\tablecomments{Rates are not corrected for extinction internal to HoII.}
\tablenum{4}
\caption{Global Star Formation Rates for Holmberg II \label{Table 4}}
\end{table*}

\section{Radial Dependence of Star Formation}

Surface brightness profiles are derived using magnitudes corrected for Galactic
foreground extinction in elliptical annuli 16$\arcsec$ wide.
The optical profiles are created using images with
the foreground star component masked; the brightest of these stars
is given the average value from their surroundings. To effectively
compare the behavior of surface brightness profiles of
irregular-type galaxies, the choice of aperture center is important.
As with the integrated magnitudes, the dynamical center is chosen as the
aperture center instead of the optical center or brightest feature
in order to make a uniform physical comparison with other galaxies.

\subsection{FUV Surface Brightness Profiles}

The patchy distribution of FUV light shown in the UIT image
is represented in the surface brightness profiles by pronounced
inflections at various radii. The FUV surface brightness profile, 
shown in Figure 3, has a double peaked shape with a
central depression. At the aperture center, the average
surface brightness is $\mu_{FUV}$=21.6 mag arcsec$^{-2}\pm0.11$ while
at the first peak, 0$\farcm9$ from the center, the average
value is $\mu_{FUV}$=21.2 mag arcsec$^{-2}\pm0.03$. Surface brightness maps
of the galaxy indicate that the highest levels of activity
lie in the region responsible for this peak; the brightest knot
in the center of the star forming arc has the highest levels,
$\mu_{FUV}\sim$18.9 mag arcsec$^{-2}$ in a resolution element (5$\arcsec$).
Following the peak, the profile exhibits somewhat exponential
behavior until r$\sim$2\arcmin,  where it takes a  constant value of
$\mu_{FUV}\sim$22.4 mag arcsec$^{-2}$. The flattening of the profile
is due to flux from the group of regions on the eastern side of
the central star forming arc. A second, yet relatively small peak
occurs at r$\sim$2.7\arcmin, after which it falls off sharply.
The second peak can be accounted for by the diffuse patches of
emission in the northern part of the galaxy.

\begin{figure}
\plotone{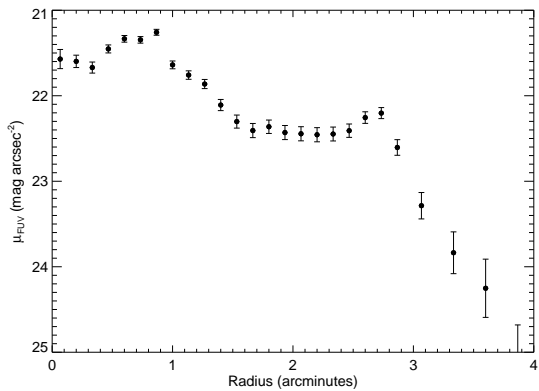}
\caption{The FUV surface brightness profile of Holmberg II. \label{Fig 3}}
\end{figure}

The sample of FUV surface brightness profiles presented by Stewart (1998) offers 
little uniformity with regard to identifying characteristics for late-type low 
surface brightness systems. It is  a relatively heterogeneous group which
does not follow either a characteristic R$^{1/4}$ law or a pure exponential.
However, some general trends do emerge which 
could have important implications in interpreting optical observations
of high redshift galaxies where the observed light is
the redshifted UV continuum. These characteristics are also exhibited by the 
HoII profile: generally flat with inflections associated with prominent star 
forming complexes, healthy activity far from the galaxy center, and non-centralized
organization of star formation. As compared to FUV profiles of other galaxies observed 
by the UIT, the HoII profile has characteristics somewhat similar to those of IBm-type 
galaxy NGC~4449 (Fanelli \ea 1997a). NGC~4449 has an exponential FUV surface brightness 
profile with inflections due to blue knots (Hill \ea 1998). The profile does not have a 
central light depression, but this may be due to the choice of the aperture center.
Although steeper than the HoII profile, the NGC 4449 profile
shows similar characteristic inflections due to star formation at large radii
from the center. 

\subsection{Optical Surface Brightness Profiles}

In contrast to the lack of FUV profiles, there have been numerous studies
characterizing the optical light profiles of low surface brightness galaxies
(Karachentseva \ea 1996; Patterson \& Thuan 1996; Vader \& Chaboyer 1994).
The B-band profile of HoII, shown in Figure 4, can be characterized as
exponential with a central depression. This behavior is typical
of late-type low surface brightness galaxies. The underlying exponential
component indicates the presence of an optical disk superimposed on
bright star forming knots while the central trough demonstrates that
star formation is not centrally organized around a nucleus.

A rough estimate of the central surface brightness of the HoII B-band profile is made
by fitting it with a pure exponential which is 
extrapolated to the center of the galaxy. Corrected for Galactic foreground extinction,
$\mu_{\rm B}$(0)=22.3$\pm0.02$ mag arcsec$^{-2}$. 
The range in central surface brightness of the galaxies in the
Karachentseva \ea (1996) sample is
22.2 $\lesssim\mu_{\rm B}$(0)$\lesssim$ 24 mag arcsec$^{-2}$ compared to the
canonical value of 21.65 $\pm0.30$ mag arcsec$^{-2}$ for disk galaxies (Freeman 1970).
Holmberg II falls at the brighter end of the observed range for low 
surface brightness irregular galaxies. Other studies report an average 
$\langle\mu_{\rm B}$(0)$\rangle$=23.5$\pm0.2$ (R\"{o}nnback and Bergvall 1994)
and $\langle\mu_{\rm B}$(0)$\rangle$=23.5$\pm1.6$ mag arcsec$^{-2}$ (Vader \& Chaboyer 1994).

\begin{figure}
\plotone{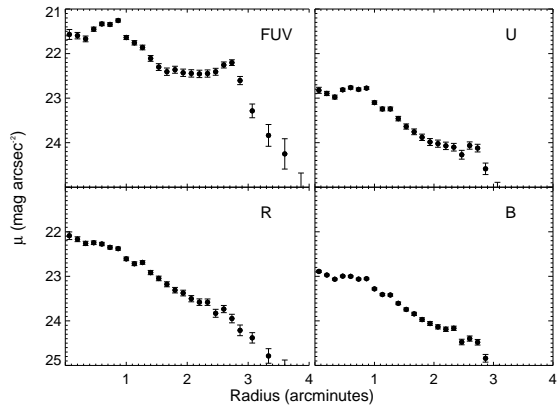}
\caption{Optical and FUV surface brightness profiles of Holmberg II. \label{Fig 4}}
\end{figure}

\subsection{Radial Continuity of Star Formation}
The general lack of correlation between star formation properties and 
other global properties such as abundance or 
global gas parameters in irregular galaxies suggests that local rather than
global conditions are important in regulating star formation. For example, Hunter, 
Elmegreen, \& Baker (1998) do not find a correlation between 
the star formation and the ratio of the HI surface density to the 
critical density for gas instabilities in a sample of dwarf irregular galaxies, while 
the stellar surface brightness does correlate with this ratio. 
These results suggest that stellar energy provides feedback for 
star formation to a certain degree. 
Studies of the radial dependence of the
optical and \ha emission also suggest that local regulation of star 
formation may be important in both irregular (Hunter \& Gallagher 1985) and spiral galaxies 
(Ryder \& Dopita 1994).
The FUV bandpass gives some insight into the radial dependence of the 
recent star formation activity, yet over longer timescales than the \ha emission allows.
Figure 4 displays the FUV, U, B, and R-band profiles of HoII on the
same linear scale. Beginning with the R-band profile and moving counterclockwise around 
the figure,
the profiles become less smooth as the wavelength
of the profile decreases. This is expected due to the contribution
from recent star formation, while at increasingly 
longer wavelengths, the light is more complicated measure of the entire 
history of star formation. The general shape of each profile is same and 
the radii at which increased star formation activity occurs remains relatively 
constant over the timescale which these profiles track star formation 
activity, roughly a few Gyrs. Figure 4 suggests that HoII is currently 
forming stars at the same radii as it has in the past, supporting the supposition 
that star formation in irregular galaxies is a local process and that 
local conditions such as gas distributions and feedback from massive stars 
may play a significant role in the star formation process. 

\begin{figure}
\plotone{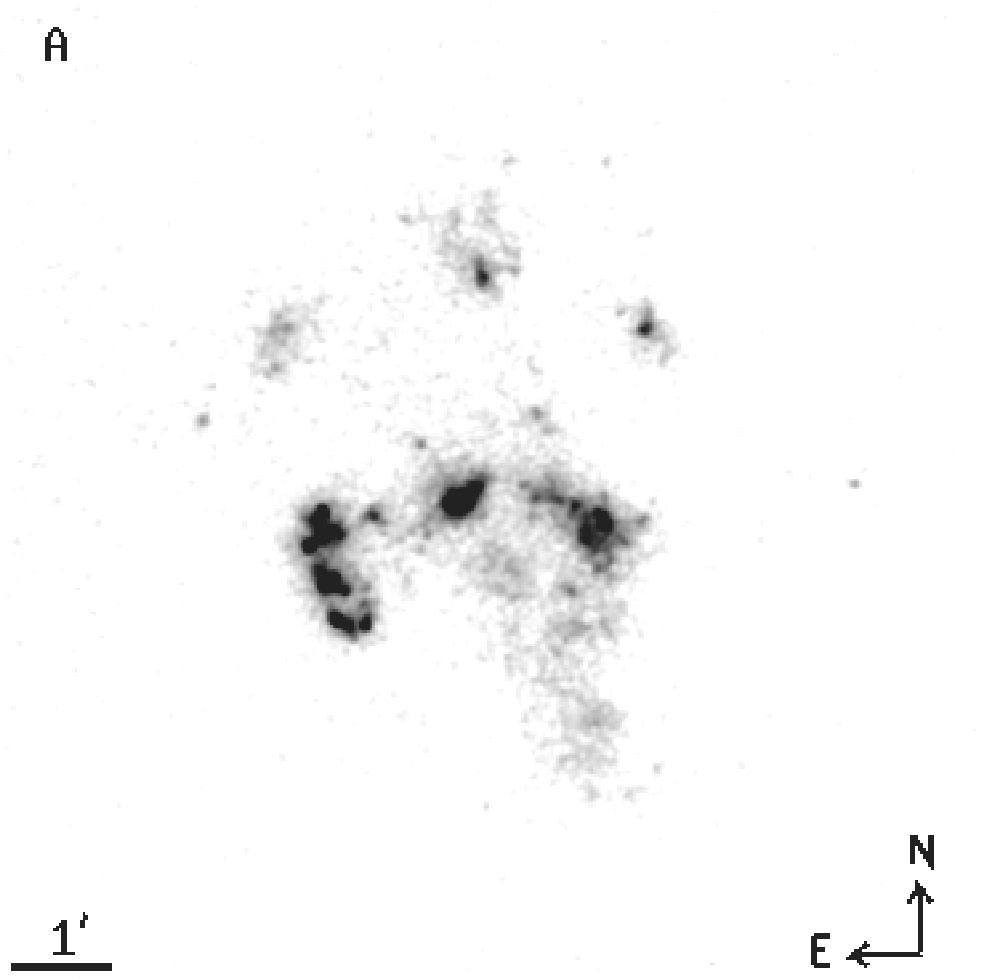}\\
\plotone{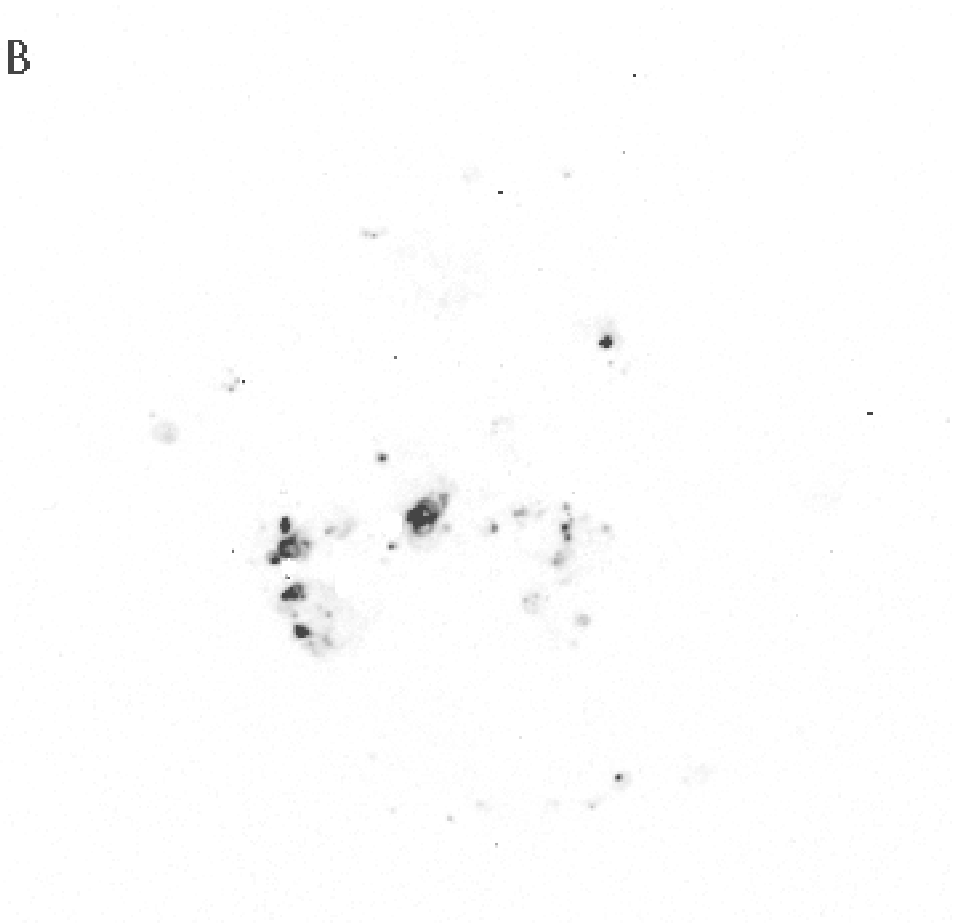}
\caption{Registered FUV (A, top) and H$\alpha$ (B, bottom) images of Holmberg II.
\label{Fig 6}}
\end{figure}

\section{Properties of Star Forming Regions}
Star forming properties are examined in
further detail by characterizing the individual star forming complexes.
Star formation for individual regions is 
quantified by means of photometry in the observed bandpassses, 
used to derive a characteristic age and internal extinction for each region. 
The physical property upon which the age derivation is based is the 
different timescale of decay of the \ha and FUV flux from an evolving cluster.
The \ha emission from an evolving cluster 
peaks at roughly 1-2 Myr and falls off sharply. The FUV emission coming 
directly from the young massive stars in an evolving cluster peaks at roughly 5 Myr 
and decays on a much longer timescale than the \ha emission (see O'Connell 1997, Fig. 1).

\begin{figure}
\plotone{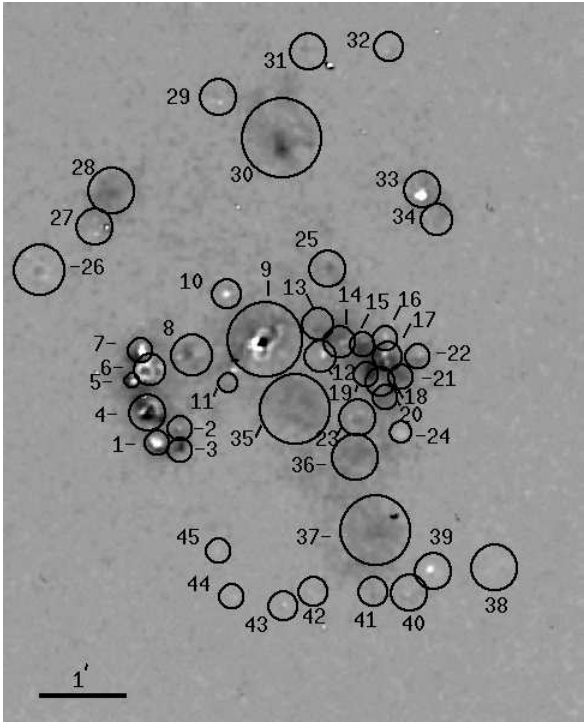}
\caption{Holmberg II star formation regions on a FUV, \ha difference image; areas 
with \ha and no FUV emission are white and areas with FUV and no \ha emission are black. 
\label{Fig 6}}
\end{figure}

\subsection{Photometry of Star Forming Regions}

Star formation regions are defined by selecting
regions of associated FUV and \ha flux. 
Since both the FUV and \ha emission define a star forming region
in this context, the ages provide a snapshot of the pattern of star 
formation over a 100 Myr timescale. This is an unconventional  
approach in defining the spatial distribution of young star formation 
which is normally described by \ha observations alone. 
The registered FUV and continuum
subtracted \ha images of HoII are compared in 
Figure 5 to highlight the differences between the spatial 
patterns of current and recent star formation activity.
\notetoeditor{Fig 5a (left) and Fig 5b (right) to be displayed adjacently to 
one another in Figure 5 with a small gap between them.}

The choice of region boundary
is facilitated by creating a FUV and \ha difference image, which 
reveals areas of associated emission. Circular apertures are 
drawn so as to include HII regions defined 
in previous studies (Hodge \ea 1994) and approximately all FUV emission above a limiting
surface brightness of 21.5 mag arcsec$^{-2}$. 
Using these criteria, 45 regions are identified. 
The apertures are drawn on a FUV, \ha difference image in Figure 6.
Some regions are defined by virtue of meeting this criteria in only one of the two 
bandpasses. The choice of aperture size is straightforward when associated 
emission is compact and coincident. Ideally, an aperture would contain only a 
coeval association and its immediate HII regions. However, the resolution of the image 
limits the ability to select a single generation of star formation. 
The effect which sampling slightly different ages in a single aperture has 
on the subsequent derivations is discussed below.

The FUV, B, and continuum subtracted \ha flux are measured for 
each region. A correction for the background light internal to the galaxy is made 
for the B-band photometry by averaging the values immediately surrounding
a region. Since this quantity may be variable for some regions,
especially the larger ones, this estimate is the greatest source
of error in the subsequent derivations. Two time dependent quantities, the FUV$-$B color
and \lograt, 
are derived for each region after correction for Galactic
foreground extinction.  The latter quantity is the logarithmic ratio 
of the number of Lyman continuum photons, converted from \ha flux assuming 
Case B recombination, to FUV flux.  The observed FUV flux, FUV$-$B color, 
and \lograt are listed in Table 5 along with 
the aperture sizes.

%
%----------TABLE 5-----------------------------------------------
%
\begin{deluxetable}{lcrcrcll}
\small
\tablecaption{Derived Properties of Star Forming Regions \label{Table 5}}
\tablewidth{0pt}
\tablenum{5}
\tablehead{
\colhead{Region}
&\colhead{Aperture} 
&\colhead{FUV}
&\colhead{FUV$-$B} 
&\colhead{O5 V}
&\colhead{Age}
&\colhead{E(B$-$V)$_{\rm i}$}
&\colhead{A$_{\rm FUV}$\tablenotemark{c}}\\
\colhead{\#}
&\colhead{($\arcsec$)} 
&\colhead{Flux\tablenotemark{a}}
& 
&\colhead{Stars\tablenotemark{b}} 
&\colhead{Group}
&
& 
}
\startdata
    1   & 16.0  &       63.8 &$-$3.18 &     24  &  1  &  0.07  & 0.63\nl
    2   & 16.0  &       41.4 &$-$3.22 &     11  &  2  &  0.03  & 0.27\nl
    3   & 16.0  &       31.3 &$-$3.45 &      7  &  2  &  0.0   & 0.0 \nl
    4   & 25.6  &      168.2 &$-$3.27 &     50  &  2  &  0.04  & 0.36\nl
    5   & 9.6   &       30.3 &$-$3.34 &     10  &  1  &  0.05  & 0.45\nl
    6   & 22.4  &      122.9 &$-$3.19 &     47  &  1  &  0.07  & 0.63\nl
    7   & 16.0  &       62.7 &$-$2.97 &     28  &  1  &  0.09  & 0.81\nl
    8   & 28.8  &       67.5 &$-$3.25 &     17  &  2  &  0.02  & 0.18\nl
    9   & 51.2  &      414.4 &$-$3.15 &    172  &  1  &  0.08  & 0.72\nl
    10  & 19.2  &       19.9 &$-$2.72 &     11  &  2  &  0.12  & 1.08\nl
    11  & 12.8  &        9.1 &$-$3.04 &      2  &  3  &  0.02  & 0.18\nl
    12  & 22.4  &       37.3 &$-$2.97 &     12  &  3  &  0.05  & 0.45\nl
    13  & 22.4  &       43.1 &$-$2.53 &     18  &  4  &  0.08  & 0.72\nl
    14  & 22.4  &       67.9 &$-$2.92 &     26  &  3  &  0.07  & 0.63\nl
    15  & 16.0  &       40.9 &$-$3.09 &      9  &  3  &  0.01  & 0.09\nl
    16  & 16.0  &       22.0 &$-$3.27 &      7  &  2  &  0.04  & 0.36\nl
    17  & 19.2  &       74.0 &$-$3.04 &     22  &  2  &  0.04  & 0.36\nl
    18  & 19.2  &       67.8 &$-$2.99 &     19  &  3  &  0.03  & 0.27\nl
    19  & 16.0  &       50.0 &$-$3.08 &     11  &  3  &  0.0   & 0.0 \nl
    20  & 16.0  &       33.4 &$-$3.18 &      7  &  3  &  0.0   & 0.0 \nl
    21  & 16.0  &       33.0 &$-$2.82 &      8  &  4  &  0.01  & 0.09\nl
    22  & 16.0  &       18.9 &$-$2.72 &      9  &  3  &  0.09  & 0.81\nl
    23  & 25.6  &       47.5 &$-$3.27 &     11  &  2  &  0.01  & 0.09\nl
    24  & 15.6  &       10.1 &$-$3.43 &      4  &  1  &  0.07  & 0.63\nl
    25  & 25.6  &        9.6 &$-$2.28 &     10  &  2  &  0.19  & 1.71\nl
    26  & 35.2  &       25.6 &$-$2.75 &     16  &  2  &  0.13  & 1.17\nl
    27  & 25.6  &       33.7 &$-$3.18 &      7  &  2  &  0.0   & 0.0 \nl
    28  & 32.0  &       61.2 &$-$2.89 &     14  &  4  &  0.01  & 0.09\nl
    29  & 25.6  &       17.9 &$-$2.73 &      8  &  3  &  0.09  & 0.81\nl 
    30  & 54.4  &      174.6 &$-$2.89 &     44  &  4  &  0.02  & 0.18\nl
    31  & 25.6  &       12.7 &$-$2.55 &      8  &  3  &  0.13  & 1.17\nl
    32  & 19.2  &        4.6 &$-$2.47 &      3  &  2  &  0.15  & 1.35\nl
    33  & 25.6  &       55.4 &$-$3.04 &     29  &  1  &  0.11  & 0.99\nl
    34  & 22.4  &       22.2 &$-$2.99 &      6  &  3  &  0.03  & 0.27\nl
    35  & 47.6  &      138.3 &$-$2.60 &     49  &  4  &  0.06  & 0.54\nl
    36  & 32.0  &       69.6 &$-$2.60 &     21  &  4  &  0.04  & 0.36\nl
    37  & 48.0  &      115.3 &$-$2.92 &     25  &  4  &  0.0   & 0.0 \nl
    38  & 32.0  &        8.7 &$-$2.42 &     10  &  2  &  0.20  & 1.80\nl
    39  & 25.6  &        9.5 &$-$2.62 &     10  &  1  &  0.19  & 1.71\nl
    40  & 25.6  &       16.2 &$-$3.12 &      5  &  2  &  0.05  & 0.45\nl
    41  & 19.2  &       13.1 &$-$3.33 &      3  &  2  &  0.0   & 0.0 \nl
    42  & 19.2  &        6.6 &$-$2.71 &      4  &  2  &  0.13  & 1.17\nl
    43  & 19.2  &        4.3 &$-$3.16 &      2  &  1  &  0.06  & 0.54\nl
    44  & 16.0  &        2.5 &$-$2.40 &      2  &  2  &  0.17  & 1.53\nl
    45  & 16.0  &        2.7 &$-$3.13 &      1  &  1  &  0.09  & 0.81\nl
\enddata
\tablenotetext{a}{FUV flux in units of 10$^{-16}$\flx, corrected for Galactic foreground
extinction only.}
\tablenotetext{b}{Equivalent number of O5 V stars at assumed galaxy distance,
derived from internal extinction corrected FUV flux.}
\tablenotetext{c}{Internal extinction derived via LMC redenning law.}
\end{deluxetable}

\subsection{Derived Ages and Extinctions}

The two time dependent parameters,  FUV$-$B color
and log (N$_{\rm Lyc}$/L$_{\rm FUV})$, are compared to model values allowing a simultaneous 
determination of the characteristic age and internal extinction of each region. 
 This method is used to estimate the ages of 
H II regions in M81 (Hill \ea 1995) and NGC 4449 (Hill \ea 1998) where it
is described in detail.
A single generation instantaneous burst (IB) model, assuming the stellar 
evolutionary tracks of the Geneva group (Schaerer \ea 1993) and 
the stellar atmosphere models of Kurucz (1992), is used. The model 
estimates L$_{\rm FUV}$, N$_{\rm Lyc}$, and L$_{\rm B}$ for
an evolving cluster over 100 Myr with a 0.5 Myr timestep. These parameters are
then integrated
over the each of the observed bandpasses. Several assumptions
are made to simulate the environment in HoII.
A mass range of 1 $-$ 100 \msol~ for model stars is assumed.
Under the assumption that regions 
are forming O-type stars, a relatively flat slope for the 
model IMF is chosen. A slope of $-$1.08, derived by Hill \ea (1994) 
for Lucke-Hodge regions in the vicinity of the 30 Dor region in the
LMC, is used. However, the subsequent age derivations are not very
sensitive to differences in the IMF. For example, using a  
model with a slope of $-$1.0 and one with $-$2.0 only results in small differences in 
the internal extinction estimates. At maximum, an underestimation of the extinction
by E(B$-$V)=0.01 for stars 3.5 $-$ 10 Myr would result. 
However, differences in the assumed cluster metallicity show substantial  
differences in the shape of the model relationship. An LMC-like metallicity,
Z/Z$_{\odot}=0.4$, is assumed.

The observed values of FUV$-$B and log (N$_{\rm Lyc}$/L$_{\rm FUV})$ 
are corrected 
with an array of E(B$-$V)'s assuming a LMC curve, 
chosen under the assumption that the 
shape of the curve correlates with metallicity (Verter \& Rickard 1998). 
In the FUV, A$_{\rm FUV}$/E(B$-$V)=9.04 for the LMC extinction curve 
(Hill \ea 1997). The array of 
corrected observations for each region are compared to the model plot.
Since the corrected data points are tied to E(B-V) values and the model is
age dependent, the point at which a corrected value agrees with the model 
curve gives both an age and internal extinction estimate.  
To illustrate
this method, the LMC model for t= 0 $-$ 50 Myr is shown in Figure 7
with several regions. The region number represents the 
uncorrected data point; the dashed line connecting it to the model 
represents the array of corrected values. The length of the 
dashed line is a measure of the internal extinction and the point at which
it crosses the model is the approximate age of the cluster. 

The age obtained using this method should be treated as the mean age of stars in the 
aperture due to the fact that a single generation model is used to interpret flux 
from what is potentially a mix of populations of slightly different ages. To compensate for
this uncertainty, regions are classified into four age groups. These groups result from both 
the shape of the model and how the observations are naturally grouped with 
one another. This range is sufficient to 
establish the relative ages of regions for large scale comparisons and to identify star formation 
patterns. 
The age groups are shown in Figure 7 by the alternating
dotted and solid lines of the model: group 1 (first dotted line) = 0 $-$ 3.5 Myr,
group 2 (first solid line) = 3.5 $-$ 4.5 Myr, group 3 (next dotted line) =
4.5 $-$ 6.3 Myr; group 4 (last solid line) $\geq$ 6.3 Myr. When comparing regions, the 
actual age cutoffs for each group should be ignored and groups should 
be thought to consist of relatively very young, young, intermediate age,
or older star forming regions for groups 1 $-$ 4, respectively. The derived age
groups and internal extinctions are listed for each region in 
Table 5.

\begin{figure}
\plotone{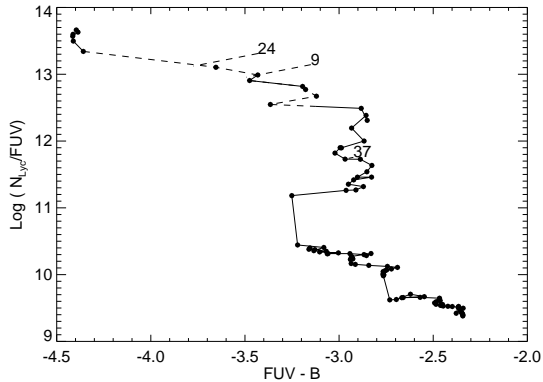}
\caption{Cluster model with three Holmberg II data points. The uncorrected
data points are represented by the region number. The age groups are shown by the alternating
dotted and solid lines of the model: group 1 (first dotted line) = 0 $-$ 3.5 Myr,
group 2 (first solid line) = 3.5 $-$ 4.5 Myr, group 3 (next dotted line) =
4.5 $-$ 6.3 Myr; group 4 (last solid line) $\geq$ 6.3 Myr. \label{Fig 7}}
\end{figure}

The ages of star formation regions are plotted on the FUV image 
to reveal the spatial distribution of star formation as a function
of age in Figure 8. The ages are identified by a symbol, 
representing the mean age of stars 
within the circular aperture of like size. 
Two locations in the galaxy generally contain the youngest regions,
the eastern side of the central arc and the linear group at the southern
part of the galaxy. Using Figure 6 to reference 
the region numbers, these are region numbers 38 $-$ 45 and 1 $-$ 9,
respectively. However, the young regions are not restricted to these
two areas. The youngest derived age, $\sim 2.4$ Myr, is located
on the western side of the galaxy (region 24) adjacent to the oldest 
group of regions. Being a strong, compact \ha source and lacking
FUV emission, the young age of this region is not surprising while
its location seems somewhat peculiar. 
The older regions in the galaxy are mostly the larger, diffuse regions
in areas of less concentrated star formation (regions 30, 35 $-$ 37). 
The oldest region, region 37, has a derived age of $\gtrsim 10$ Myr.
It is most likely much older than this, but difficulties in measuring 
the \ha flux at this level do not allow the limit to be constrained 
further. The origin of diffuse FUV emission found in a galaxy is
sometimes hard to determine. Possible sources include scattering 
from dust grains and light from a spatially unresolved population
of hot stars. In a low metallicity galaxy with low 
global extinction, scattering of UV light is not problematic and 
allows direct imaging of high-mass star formation. Therefore, the 
diffuse light in the larger apertures is most likely a spatially
dispersed population of B stars formed over the past few 100 Myr.

\begin{figure}
\plotone{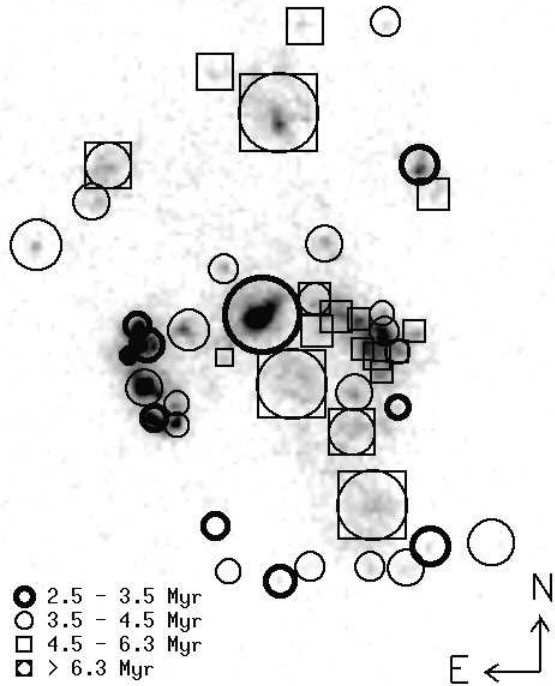}
\caption{Derived ages of Holmberg II star forming regions plotted on the FUV image. \label{Fig 8}}
\end{figure}

The derived internal extinctions, A$_{\rm FUV}$, in Table 5 show 
several trends when compared to their respective age groups. 
Age groups 1 $-$ 3 show a full range in extinction, A$_{\rm FUV} \sim 0 - 2$
while regions in group 4 have substantially lower extinctions, generally
A$_{\rm FUV} \lesssim 0.3$. Extinction estimates for individual star forming 
regions derived from H$_{\alpha}$/H$_{\beta}$
line ratios for dwarf galaxies, including 4 regions in HoII, also fall in this
range (Hunter \& Gallagher 1985). 
Extinction proves to be quite patchy when comparing the 
spatial distribution of extinction levels. This distribution can be reconciled 
with the shell-like HI morphology of the galaxy (discussed in \S7).  The highest extinction levels
are found in the dense HI ridges and the lowest levels in the HI voids. 
There is often not a smooth transition between extinction levels of adjacent 
regions. The distribution of internal extinction levels in HoII illustrates the 
problematic nature of using a global correction in irregular galaxies, especially 
those with HI shell structures.

\subsection{Number of O Stars}

To further quantify the FUV fluxes in individual star forming regions, the 
number of equivalent early-O type main sequence stars is derived from the 
internal extinction corrected FUV flux of each region and listed in Table 5. 
This exercise allows a 
comparison between the relative stellar content of regions to the first order. 
The derivation is
based on the M$_{\rm FUV}$ for normal stars in the UIT bandpasses from the 
Fanelli \ea (1992) stellar library.  To derive the equivalent number
of stars per a given region, the expected flux of a single O5 V
star is scaled to the galaxy's distance and compared to the corrected FUV flux. 
A single O5 V star is M$_{\rm FUV}$(B1)=$-10.2$; its unextincted magnitude 
at the distance of HoII is m$_{\rm FUV}$(B1)=17.2.

An equivalent number of 172 early-O type stars is estimated for the most
energetic complex in HoII, region 9. This number is comparable to,
but slightly less than, the number generally accepted to be present in the 
central (1$\arcmin\times1\arcmin$) 30 Dor complex in the LMC, $\sim$ 230 (Vacca \ea 1995). 
Using FUV 
data, Hill \ea (1997) estimate that the brightest 
regions in Sbc spiral M51, Rand 308 and 78, have an 
equivalent number of 245 and 242 early-O type stars, respectively.  
Holmberg II is dominated by one 
large complex which is able to produce 
a prodigious amount of star formation comparable to that in much larger galaxies. 
Even the large, FUV diffuse regions have a significant equivalent number of
early-O type stars present. For example, the diffuse FUV flux 
in region 35 is equivalent to 49 O5~V stars.    

\section{Massive Stars and the Local ISM}

Massive stars dump energy into their surrounding ISM via
ionizing photons, supernovae explosions, stellar winds, and
outflows. Combined, these processes have a profound affect 
on the surrounding region of gas. The ionizing photon 
bath alone, provided by UV photons from massive stars, can 
deposit prodigious amounts of kinetic energy into the ISM.
The ``champagne'' model (Tenorio-Tagle 1979), summarized 
by Hunter (1992), describes the 
process by which massive stars ionize their surrounding gas 
creating compact HII regions that evolve into extended objects. 
As the ionization front reaches the edge of the parental molecular 
cloud, ionization of the neutral medium commences, where by
the stars lie in a half-open cavity in the dense cloud and 
the HII region emerges as a ``blister'' on the side of the 
cloud. Numerous observationally-based studies support both the existence of champagne 
flows and the premise that expanding ionized structures are caused by localized 
energy sources (Lafon \ea 1983; Heydari-Malayeri \& Testor 1981). 
A tedious approach to 
identifying massive stars inside ionized structures can be made 
through spectroscopy. Optical observations of massive stellar populations, the 
less massive remnants of the population actually responsible for the expanding  
features, have been employed to compare the energy available from the population with 
the ionization and formation requirements of the expanding ionized structures 
(Hunter \ea 1995).  In contrast, Parker \ea (1998) use UIT observations
of the LMC and SMC to identify the hottest OB stellar members, those responsible 
for the majority of the ionizing flux, and correlate the FUV photometry of these 
stars with the \ha flux of the associated HII regions.  

\subsection{Morphology of Massive Stars and Ionized Gas}

Several examples illustrating the relationship between massive 
stars and ionized structures in various evolutionary stages are
found in HoII. Figure 9 shows registered \ha and FUV images of region 6. 
The shell in the center of the \ha image encircles a group of massive
stars present in the FUV image providing evidence of a causal relationship.
The emission on the edges of the \ha shell itself is not symmetrical; the 
emission on the left side of the shell is more prominent than on the 
right, consistent with the champagne model. Region 9 presents 
similar evidence in Figure 10 where a more complex system of shells is 
present, suggesting that a less spatially concentrated group of massive stars 
is responsible for the ionization. The FUV image identifies the 
distribution of these stars, a $\sim 400$ pc chain
decreasing in intensity from the lower right to the upper left.
The FUV luminosity of this region indicates the presence of an equivalent number 
of 172 O5 V stars; an average age of $\sim 3.3$ Myr is estimated for the entire 
complex.

\begin{figure}
\plottwo{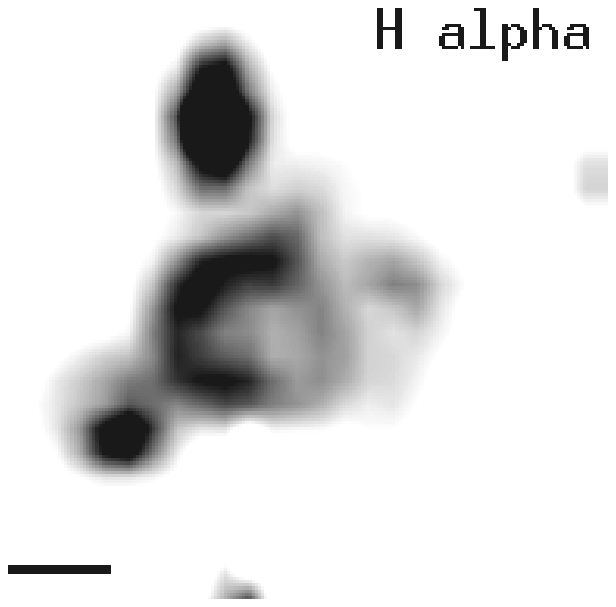}{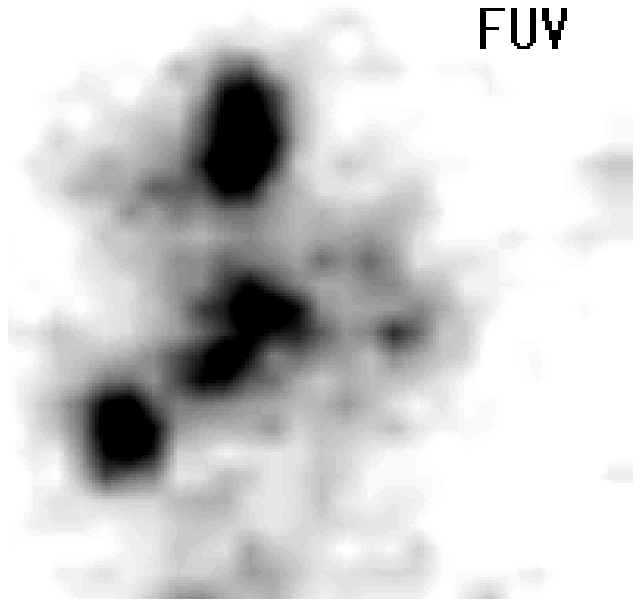}
\caption{Registered \ha and FUV images of Holmberg II, region 6. 
The dark bar in the lower left-hand corner represents 100 pc in Figures 9, 10, 12, and 13. \label{Fig 9}}
\plottwo{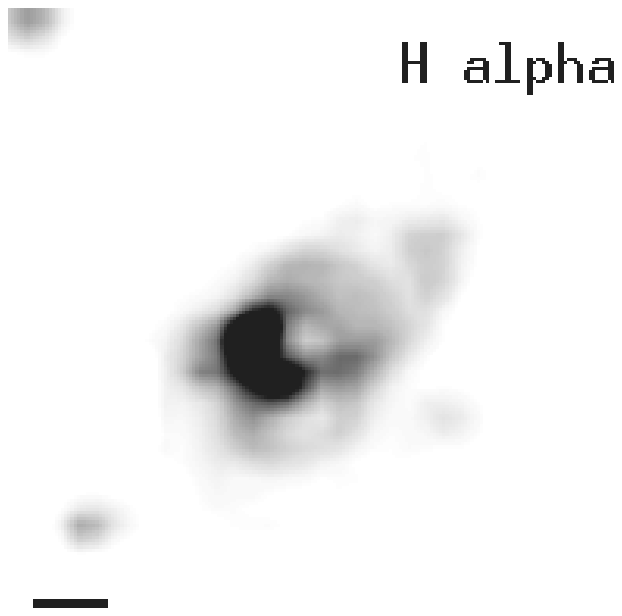}{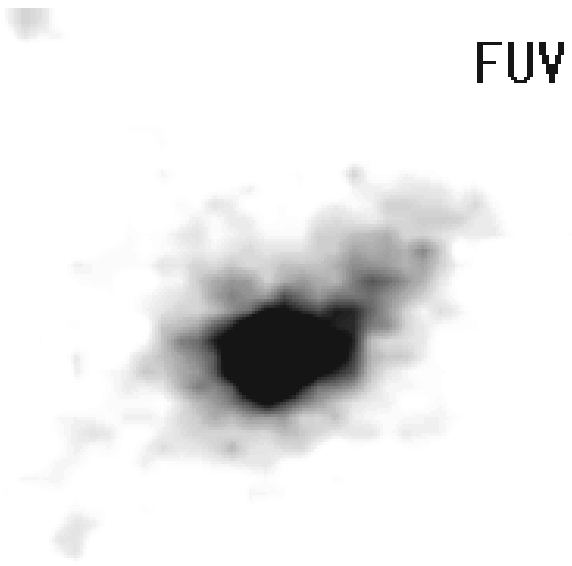}
\caption{Registered \ha and FUV images of Holmberg II, region 9.\label{Fig 10}}
\end{figure}

\subsection{Local Triggering Mechanisms}

As illustrated by Elmegreen \& Lada (1977), 
the high pressures generated by stellar winds, ionized gas, and
supernovae explosions can act as a triggering mechanism to convert stable 
configurations of gas to unstable ones. The ionization front emerging from a 
star formation complex is followed by a 
shock front generated by the supersonic flow of the ionized material. 
If formed on the edge of a molecular cloud, the successive
passage of ionization and shock fronts into a 
neighboring cloud allows unstable 
configurations of dense neutral material to accumulate to form new 
stars. These stars generate new ionization fronts and the cycle continues 
(Blaauw 1991). Massive star formation has the ability to influence the formation of new sites
of star formation creating linear sequences of stellar subgroups within OB 
associations, potentially allowing self-propagation throughout a region.

The cumulative chain reactions initiated by these local events have 
been suggested as an overall mechanism for star formation in irregular 
galaxies. First proposed by Gerola \& Seiden (1978) as 
a global mechanism for star formation, the stochastic self-propagating 
star formation model (SSPSF) describes the propagation of star formation 
between neighboring regions by means of a ``percolation'' process. 
Gerola \ea (1980) adapt the SSPSF to small galaxies with no differential 
rotation and find the model predicts characteristics which correlate 
nicely with those of dwarf galaxies. In particular, the model produces
galaxies that form stars in a bursting mode of short-lived episodes
and a SFR that varies drastically with time. While the average SFR's
are low, the galaxies have instantaneous SFR's that range from bursting
to quiescent, which could explain the diverse star formation characteristics
observed in dwarf galaxies. However, a recent survey of 110 Virgo dwarf 
irregulars indicates that star formation is generally preferential to the 
edges, mostly to one side of the galaxy, while the SSPSF model does not predict
patterns that predominately favor any particular location (Brosch, Heller, 
\& Almoznino 1998).  

\begin{figure}
\plotone{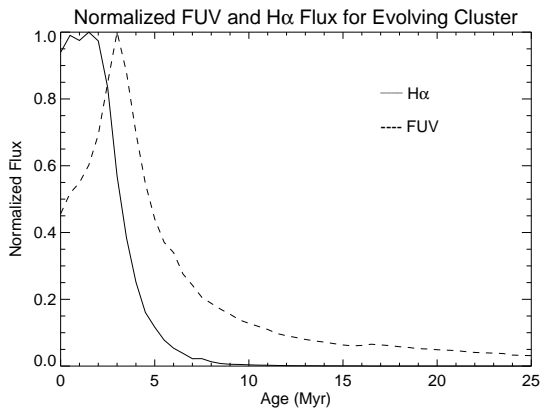}
\caption{Normalized FUV and \ha flux for an evolving star cluster.\label{Fig 11}}
\end{figure}

Sequential star formation has been documented observationally in 
galaxies of similar type as the present sample (Lortet 
\& Testor 1988; Dopita \ea 1985). 
On local scales (within the same molecular cloud), sequential star 
formation can be inferred by observations of two spatially adjacent 
regions with slight age differences. 
Comparisons of \ha and FUV emission morphologies can indicate
the relative ages of adjacent young star formation sites due to
the longer timescale of decay of the FUV emission in a given region.
The relationship between the \ha and FUV flux as a function of 
time for a given cluster is shown in Figure 11 illustrating the  
physical basis for this qualitative treatment of relative cluster 
ages. A relatively strong FUV source surrounded by compact HII regions 
is suggestive of sequential star formation using this
method. Region 23, shown in Figure 12, exemplifies this scenario.
The region, having an average age
of $\sim4.3$ Myr, consists of two compact HII regions surrounding a relatively 
strong FUV source. 
The group of regions on the western side of the central star 
forming arc (regions 15-22), shown in Figure 13, illustrates the geometry 
of sequential star formation as suggested by the SSPSF model.
Age estimates of the individual regions show a mix of age groups
2$-$4. A chain of young star
formation seen on the \ha image is located to the right
of a faint chain of FUV emission suggesting star formation is 
propagating from left to right. 

\begin{figure}
\plottwo{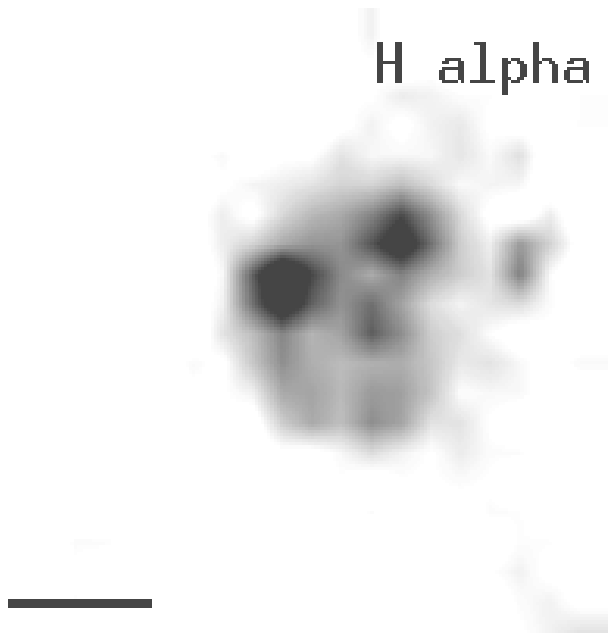}{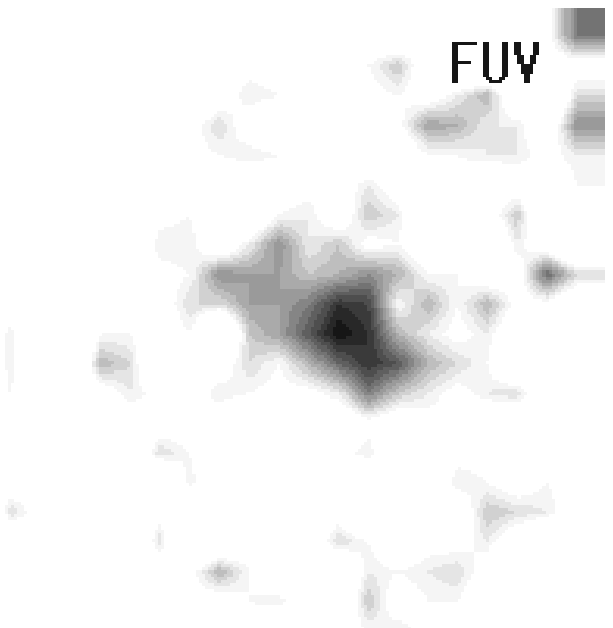}
\caption{Registered \ha and FUV images of Holmberg II, region 23.\label{Fig 12}}
\plottwo{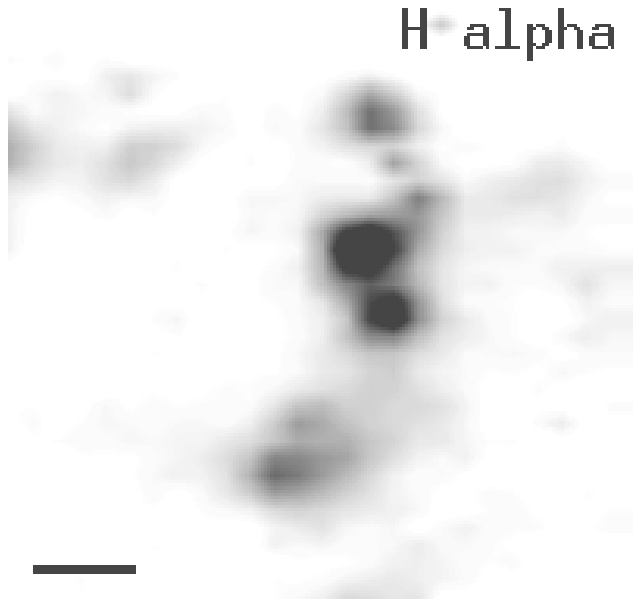}{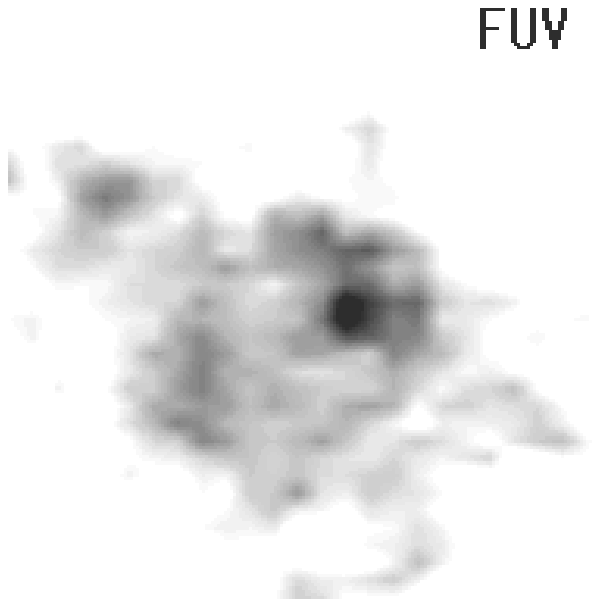}
\caption{Registered \ha and FUV images of Holmberg II, regions 15-22.\label{Fig 13}}
\end{figure}

\section{Large-Scale Triggering Mechanisms}

Massive star formation can influence 
new star formation activity on larger scales as high pressures from supernovae push gas away 
from aging OB associations. In theory, when a star forming complex has sufficient numbers of 
main sequence O and B stars, the combined pressures from stellar winds and supernovae are 
able to drive an expanding ionized shell of gas and at the same
time, create a void in the neutral gas (Tenorio-Tagle \& Bodenheimer 1988). 
Sequential star formation typically acts on only one side of a molecular cloud where as 
large-scale triggering describes 
the process by which a centralized source destroys the remainder of 
its parental molecular cloud, creating a void in the gas and a dense ridge of
swept up material carried to large distances (Weaver \ea 1977). Described by Elmegreen (1992) as the 
``collect and collapse'' method, dynamical instability of the accumulated gas collapses under 
self gravity to form new stars along the periphery of the shell. 
This scenario is a plausible explanation for the correlation between stellar 
age and proximity to the HI shell boundary in dwarf galaxy Sextans A, where the 
youngest stars are primarily found in the interior of the HI shell (Van Dyk, Puche, \& Wong 1998).
The ridge surrounding a swept-up shell may fragment leaving chains or arcs of 
new star formation, allowing a means for star formation 
to propagate on large scales. Efremov \& Elmegreen (1998) recently demonstrated 
that two arcs of clusters inside and on the rim of the superbubble LMC4 in the LMC, which 
surrounds a roughly 30 Myr-old dispersed group of supergiants and Cepheid variables,
may have 
been formed by the self-gravitational collapse of gas in swept-up pieces of rings 
or shells.  On smaller scales, feedback from
massive star formation is also evident in HI data, such as the turbulent and 
fractal nature of the local ISM shown by the HI mosaic of the LMC (Kim \ea 1998).
 
Puche \ea (1992) point out that
the observed radial expansions, energy requirements and kinematic ages 
of the holes in HoII indicate that they originate from internal,
pressure driven events. Radio continuum observations of HoII confirm that supernovae 
do lie within several HII regions (Tongue \& Westpfahl 1995). Further analysis indicates that 
the supernovae energy injection rate into the galaxy is sufficient to account for the 
global number of HI features. On the other hand, 
numerical simulations using hydrodynamic methods to simulate powerful local energy 
sources have failed to reproduce all the features in the neutral medium of HoII 
(Maschenko \& Silich 1995). Some of the HI holes in HoII, especially those located
beyond the optical disk of the galaxy, are likely formed by alternate mechanisms such 
as gravitational instability via collisional dissipation of gas particles (Byrd \& Howard 1992) 
or infall of gas clouds (van der Hulst \& Sancisi 1988).  Rhode \ea  (1999) identify a 
single high velocity cloud candidate in their HI data. However, the distribution of 
holes is too regular and present at all radii to be explained entirely by infall.
Other environmental factors, such as tidal perturbations, are not a factor due to
the fact that HoII is fairly isolated.

The closest neighbor to HoII is M81 dwarfA (Kar 52).
Given that the kinematic HI mass of HoII is a factor of 100 greater 
than that of M81 dwarfA (Melisse \& Israel 1994),
it is unlikely to have affected HoII.  Moreover,
M81 dwarfA is at a distance on the sky of roughly 27 kpc 
from HoII with a relative radial velocity of 44 km/s.
The current tidal perturbation of HoII due to this neighbor will
be very tiny compared to the internal gravitation.
Using the values of the known components of position and
velocity along with the masses, a positive two body
orbital energy is obtained.  The calculated value is less than
the actual positive value if the unknown components of position and
velocity were included. Thus, analysis of the orbital energy indicates 
that the companion is likely to be in a hyperbolic orbit.
There is not a high probability of a close encounter, especially a
prolonged one. The hyperbolic relative velocity with HoII indicates 
that an interaction might have happened $\sim$600 Myr ago, roughly three 
rotation periods within 1.8 kpc (the majority of the optical portion of the galaxy). This 
timescale is sufficient for the effects of a weak interaction 
to have dissipated via differential rotation.

Rhode \ea (1999) search for the lower-mass 
remnants of the population which in theory created the HI holes in 
several dwarf galaxies, including HoII. They 
estimate the expected number of A and F main sequence stars from the
kinematic ages of the holes and the energy required for their creation.
They do not find sufficient
numbers of these stars in the HI holes in deep optical searches to support 
the stellar wind/supernovae scenario, although the expected number is
dependant upon the choice of IMF.  In addition, Massey \ea (1996) find that 
the UV-bright stars in M33 are evolved B supergiants indicating that 
the optically brightest stars are not necessarily the ionizing sources.
Since FUV observations isolate the massive stellar component 
on timescales similar to the kinematical ages of HI holes, they can 
provide an observational ``smoking gun'' to the theoretical concept of pressure
driven shells. Instead of searching for a requisite number of lower-mass
neighbors,  a search for the progenitor stars themselves can be made. 
In addition, FUV observations can show  
sites of secondary star formation in the dense ridges, allowing insight into the possibility that 
star formation triggering via giant HI shells is a viable mechanism
for star formation. FUV observations have the potential to provide a snapshot of a  
full cycle of propagating star formation. 

\subsection{Relative FUV and HI Morphology}

The numerous voids in the HI morphology of HoII are evident on the 21cm map
presented in Figure 14; the FUV surface brightness contours are
overplotted on the image. The large spatial extent of the HI disk is also 
apparent, extending well beyond both the FUV and optical disks. The HI holes
are distributed evenly across the entire HI distribution including 
regions extending beyond the stellar disk. The largest hole is a $\sim$ 1.7 kpc diameter
void with a dense ridge in the southern part of the galaxy.

\begin{figure}
\plotone{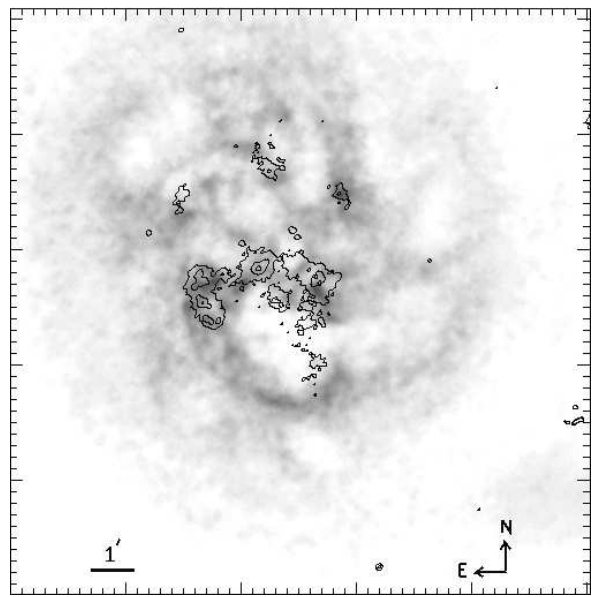}
\caption{HI image of Holmberg II overplotted with the 19, 20, and 21.8 mag arcsec$^{-2}$ FUV surface brightness contours.\label{Fig 14}}
\end{figure}

Since the expansion velocities of the HI holes were likely 
greater in the past, estimates of the kinematic 
ages using present day expansion rates yield upper limits. 
Puche \ea (1992) derive ages for the HI holes ranging from 26 $-$ 136 
Myr assuming a constant expansion rate. Given these ages, some
measure of FUV emission should be spatially coincident with the voids 
if they are caused by massive star formation. Even if only the most massive 
stars are responsible for the generation of the HI holes, their 
less massive neighbors (early B-type stars) would still be detectable
on the FUV images. However, the existence of HI holes beyond 
the stellar disk is not consistent with the supernovae/stellar wind 
scenario for their creation unless a very unusual IMF is present in 
the outer parts of the galaxy. It is likely that an alternate 
mechanism is responsible. However, this mechanism is not a viable 
feedback mechanism for new star formation due to the absence of dense rims 
and corresponding secondary sites on the periphery of these holes.

The lack of FUV emission in the centers of HI holes in the 
outer parts of HoII does not disprove the generation of
large shells by massive stars since several HI voids  
in the inner part of the gas distribution are coincident with FUV emission. In addition,
density enhancements along the rims of HI shells correspond to 
sites of FUV emission suggestive of propagating star formation.
The largest HI hole illustrates the connection between two generations
of star formation. The FUV-bright central star forming arc is aligned with 
the northern rim of this HI hole, while diffuse patches of FUV emission
lie within its boundary. Other examples of FUV emission coincident 
with HI voids are the eastern two patches in the northern 
part of the galaxy. Several of the second generation 
sites also show they are beginning to disrupt the surrounding gas. 
For example, the intense star formation on the eastern edge of the 
central arc has HI density enhancements located on either side. 
Tongue \& Westpfahl (1995) confirm the existence of supernovae 
inside star forming regions in this area (regions 
1, 4, and 6) using radio continuum observations. These supernovae
are likely responsible in part for reshaping the surrounding neutral medium.
Photodisassociation of H$_{2}$ on the interior of surfaces of the 
molecular cloud may also be responsible for the HI density enhancements
surrounding the intense young star forming regions to a certain degree. However, Tacconi \&
Young (1987) do not detect any $^{12}$CO emission above the 3$\sigma$
level in HoII. Lacking other observations to trace the molecular
component, its morphology remains a mystery. In general, evolved star forming knots are fairly
efficient in converting in converting H${_2}$ to HI (Allen \ea 1987). 

\begin{figure}
\plotone{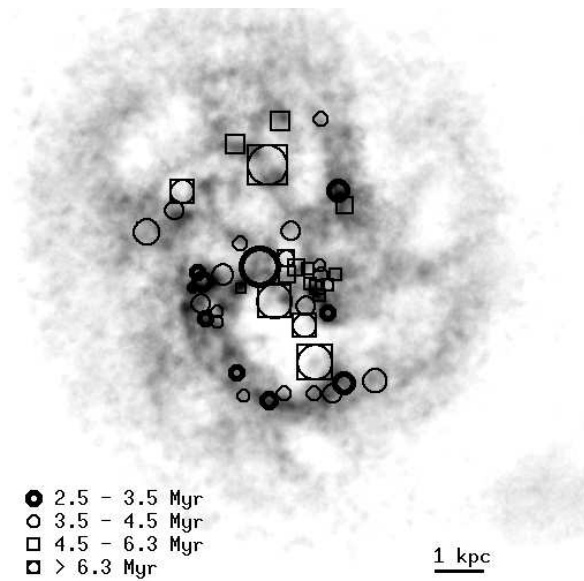}
\caption{HI image of Holmberg II shown with the derived ages of star forming regions.\label{Fig 15}}
\end{figure}

The derived ages for individual star forming regions further illustrate 
the connection between massive star formation and the HI morphology. 
The age groups for individual star formation regions
are displayed in Figure 15 on the HI map. Each of the seven regions in the 
oldest age group (group 4, $> 6.3$ Myr), lies in a HI hole or local minimum
in the HI density. The largest hole contains three large regions in 
age group 4 on the western side of its interior.  The youngest
regions (group 1, 2.5 $-$ 3.5 Myr) are each spatially coincident with HI density 
enhancements suggesting the potential for new star formation is 
high on the dense rims of the HI holes. The majority of new
star formation in the whole system is occurring on the periphery of the
largest hole. The distribution of star formation ages in general can 
be interpreted by this comparison. The spatial proximity of a very 
old region and a very young one is reconciled by the HI density distribution.

\subsection{HI Hole Energetics}
\begin{figure}
\plotone{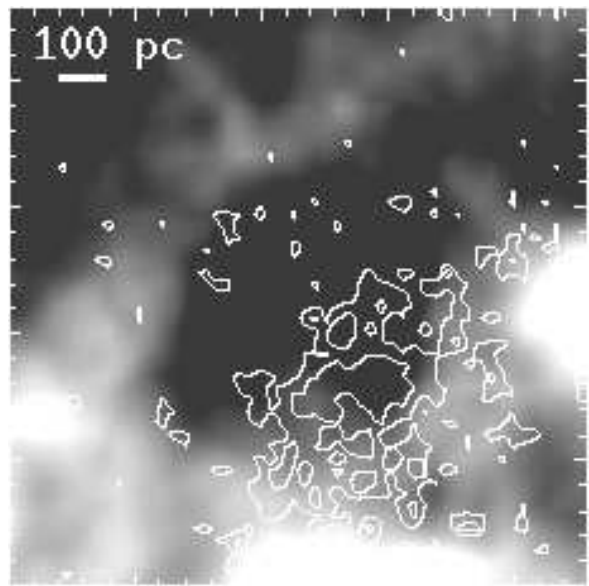}
\caption{HI image of Holmberg II region 28 overplotted with the 21.5 and 22 mag arcsec$^{-2}$ FUV surface brightness contours. Figures 16$-$18 are displayed so that the HI holes appear as dark voids and the HI density enhancements are bright.
\label{Fig 16}}
\end{figure}
\begin{figure}
\plotone{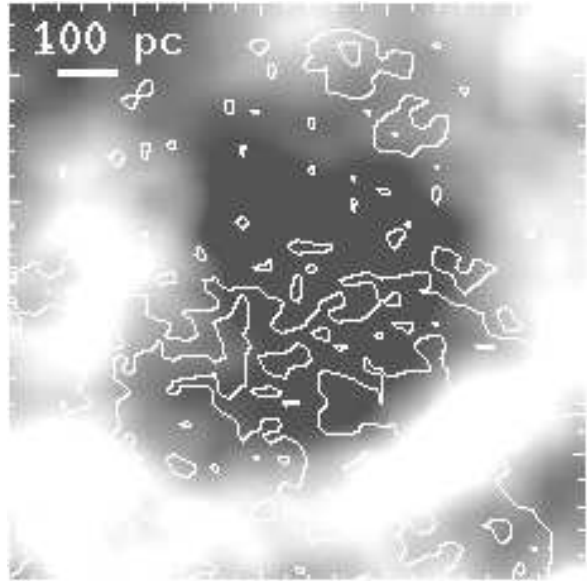}
\caption{HI image of Holmberg II region 13 overplotted with the 21 and 21.7 mag arcsec$^{-2}$ FUV surface brightness contours.\label{Fig 17}}
\end{figure}

A quantitative approach to test the observational evidence 
suggesting that massive star formation triggers holes in the neutral
medium is offered by comparing the energy delivered to the ISM 
by massive stars and the required creation energy of the HI holes. 
FUV observations of the clusters interior to the 
HI holes provide an estimate of the total mechanical energy delivered to the
ISM, assuming the derived ages and internal extinctions. Several 
of these clusters are spatially coincident with HI holes included 
in the Puche \ea (1992) study where estimates for their creation energy, E$_{\rm o}$,
are given. These hole/cluster combinations  
are listed in Table 6. Region numbers refer to the previously 
defined star formation regions and hole numbers are taken from 
Puche \ea (1992).  For reference, 
HI images showing each hole used in this comparison are presented with the 
FUV contours in Figures 16-18.
\begin{figure}
\plotone{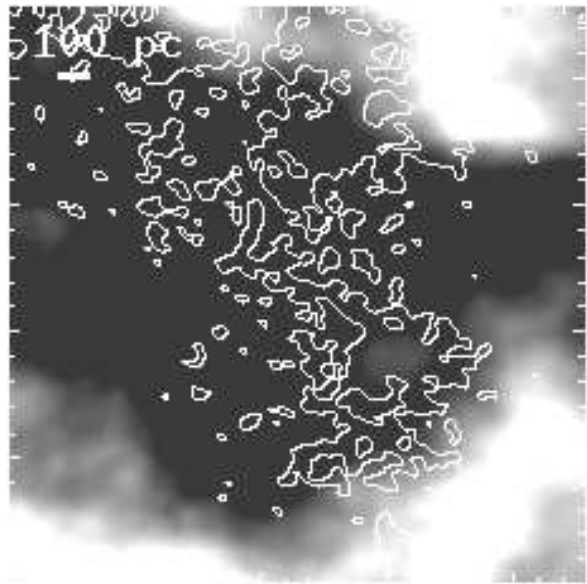}
\caption{HI image of Holmberg II regions 36 and 37 overplotted with the 19, 20.5, and 21.75 mag arcsec$^{-2}$ FUV surface brightness contours.\label{Fig 18}}
\end{figure}

The observed FUV luminosity, L$_{\rm FUV}$, for each cluster  
is estimated from the FUV flux and internal extinction correction given 
in Table 5. Cluster mass is estimated by 
comparing the observed L$_{\rm FUV}$ with the model of 
L$_{\rm FUV}$ (ergs s$^{-1}$ ~\msol$^{-1}$) for an evolving cluster at the 
given cluster age. This model is used to derive the ages and
internal extinctions for all star formation regions and 
described previously (\S 5.2). The mechanical energy deposited 
into the ISM, L$_{\rm mech}$, over a timescale equal to the age of the 
cluster can be calculated assuming a rate which is a function of cluster mass. 
Evolutionary synthesis models of populations of massive stars given
by Leitherer \& Heckman (1995) provide estimates of the deposition 
rates from stellar winds and supernovae 
for an instantaneous burst as a function of time and mass. From 
clusters of similar IMF and metallicity, an average rate of 
L$_{\rm mech}\sim3\times10^{34}$ ergs (s~ \msol)$^{-1}$ is assumed.
The net mechanical energy deposited 
into the ISM by supernovae and stellar winds, E$_{\rm SN}$, is
estimated assuming this rate, the total cluster mass, and the derived 
cluster age. 

%----------------TABLE 6------------------------------------
%
\begin{table*}
\begin{center}
\begin{tabular}{lccccccc}
\tableline
Reg & Hole & Age  & L$_{\rm FUV}$        &     Mass    & E$_{\rm o}$  &E$_{\rm SN}$   \\
\#  & \#   & Myr &10$^{36}$ ergs s$^{-1}$&$10^{3}$\msol& 10$^{50}$ergs& 10$^{50}$ergs  \\
\tableline\tableline
28   &48 &8.5 &7.4  &7.57 &470.3 &608.8  \\
13	&23 &8.5 &9.3  &9.53 &41.1  &766.4  \\
36,37&21 &10.0&23.6 &32.10&2025  &3036.9  \\
\tableline
\end{tabular}
\end{center}
\tablecomments{Region numbers refer to clusters defined in Table 5. Data for the HI holes
encircling a given cluster are taken from Puche \ea (1992) including hole numbers and
creation energy, E$_{\rm o}$.}
\tablenum{6}
\caption{Derived Properties of Clusters in HI holes}\label{Table 6}
\end{table*}

The derived properties of the clusters spatially coincident to HI
holes are given in Table 6. In each case, E$_{\rm SN}$ $>$ E$_{\rm o}$
indicating that to the first order, the energy provided by massive stars is sufficient to 
create the HI holes that encircle them. The largest hole 
(\# 21) actually contains two star forming regions which are added 
together to estimate the energy input. All the star forming 
regions  interior to HI holes are in the oldest age group 
(group 4, $>$ 6.3 Myr). The derived ages in Table 5 for these regions
are upper limits in the sense that these regions are likely much
older. (The technique to derive the ages depends upon the presence
of \ha emission and is not reliable for absolute ages $>$6.3 Myr.)
For these regions, E$_{\rm SN}$ is a lower limit and would increase 
linearly with the true cluster age. Region 13 has significantly
more energy than the creation energy of its associated HI hole. It is 
possible that this region is also responsible for an adjacent hole (\#~ 22),
which would increase E$_{\rm o}$ by 35 $\times10^{50}$ ergs. 
These comparisons suggest that massive stars do provide sufficient 
mechanical energy to account for the observed properties 
of associated HI holes in HoII.

\begin{figure}
\plotone{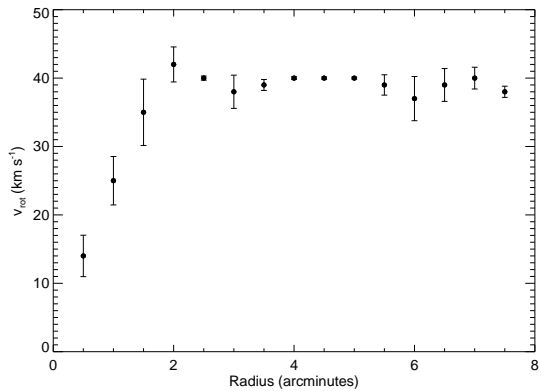}
\caption{Holmberg II rotation curve taken from Puche \ea (1992).\label{Fig 19}}
\end{figure}

\subsection{Characteristic Differences in the ISM}

The question remains: if massive star formation can provide 
sufficient energy to generate holes in the surrounding neutral medium,
why are there not HI holes around every OB association? The
answer may be related to global properties of the ISM. Characteristic
differences in the kinematical properties of the ISM can be traced to 
differences in the star formation mechanisms. 

Solid body rotation of the gas disk is a general property of dwarf galaxies
which makes them appealing for star formation studies. The lack of
shear in the absence of differential rotation allows features in the 
HI gas to be longer-lived. The thicker gas disks and lack of 
shear in irregulars allow the growth of larger gas clouds (Hunter 1997). 
This combination
provides an excellent environment for stars to form in giant complexes.
Since larger complexes impact the surrounding ISM to a greater 
degree and HI features survive longer in regions of reduced shear, 
HI holes and associated sites of secondary star formation
are more likely to be found in 
areas of the gas disk exhibiting solid body rotation. 

\begin{figure}
\plotone{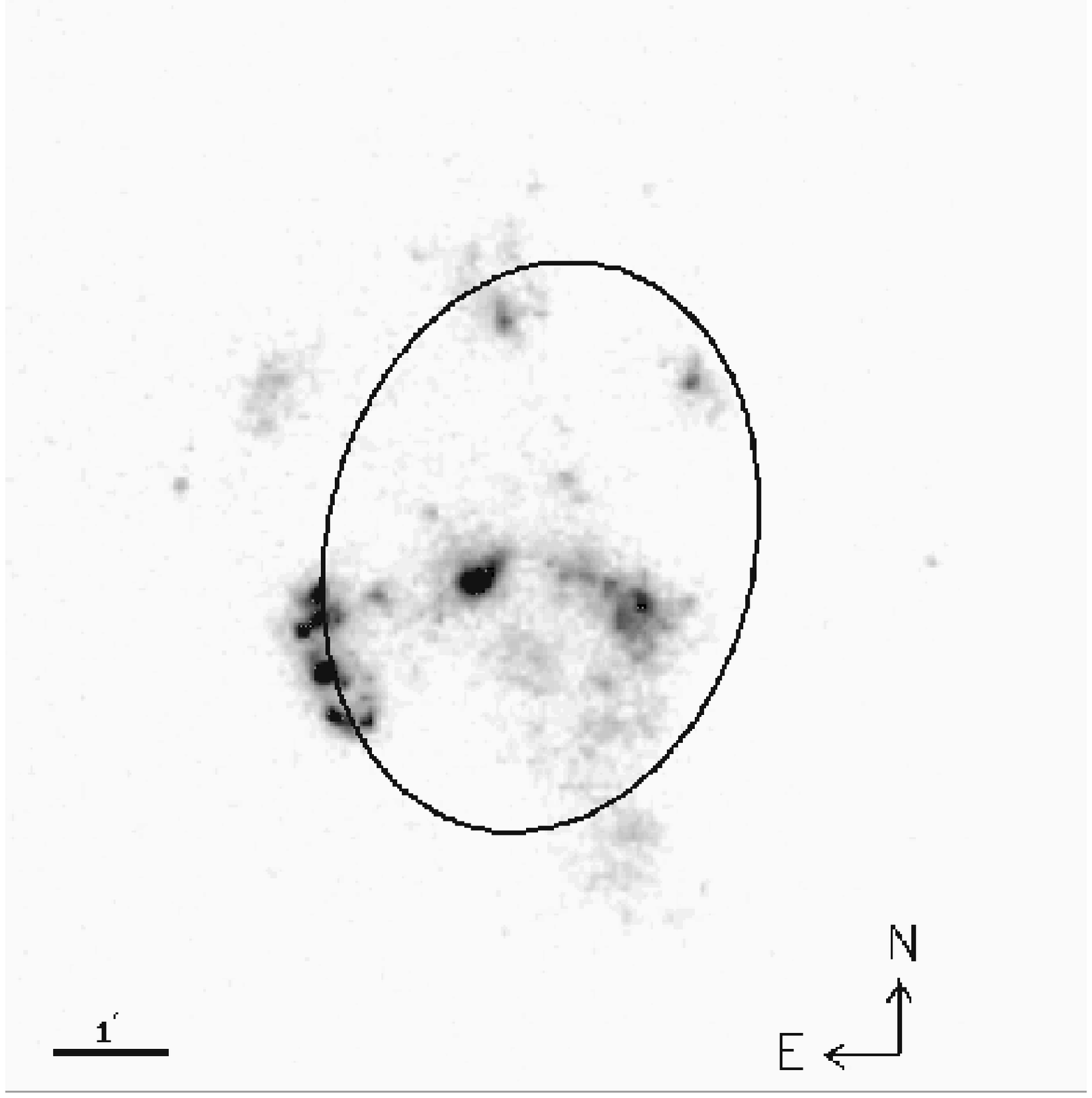}
\caption{HI image of Holmberg II shown with a 2$\arcmin$ ellipse. The interior of the ellipse represents the solid body regime of the neutral gas disk.\label{Fig 20}}
\end{figure}

The rotation curve of HoII, taken from Puche \ea (1992) and displayed in Figure 19, 
exhibits solid body rotation to a radius of r$\sim2$\arcmin~ after which it flattens
and exhibits differential rotation. An ellipse with a major axis of 2\arcmin,
reduced to the plane of HoII, is plotted on the FUV image of the 
galaxy in Figure 20. The majority of the FUV emission is contained
within the ellipse. This is also the region where examples of
large-scale star formation triggering via HI shells are found. This simple
illustration demonstrates that massive star formation in HoII can operate more
easily as a feedback mechanism in regions with uncomplicated 
global kinematical properties.

\section{Summary}

This study uses FUV, HI, and \ha observations of dwarf irregular galaxy 
HoII to trace the interaction between sites 
of massive star formation 
and the neutral and ionized components of the surrounding ISM
and to identify the means by which star formation propagates
in this intrinsically simple system. Both local and 
large-scale triggering mechanisms related to massive stars are identified 
suggesting that feedback from massive stars
is a microscopic process operating in all galaxies to a certain degree. 

The key results from the study can be summarized as follows:

(1) In comparison to its optical images, the FUV morphology of HoII
is patchy and irregular. Massive star formation does not have
a smooth spatial distribution; it is concentrated in several 
dominate complexes. The fact that 
massive star formation is not centrally organized 
has important implications for interpreting images of 
high redshift galaxies (z $\gtrsim$ 3) where  
the restframe FUV is detected using optical and IR instruments.

(2) Surface brightness profiles are used to characterize the 
radial dependence of the azimuthally averaged star formation activity. 
The FUV profile is generally flat with inflections due to prominent 
complexes. The B-band profile displays an exponential shape with 
a central light depression, characteristic of dwarf galaxies. The  
derived central surface brightness, $\mu_{\rm B}$(0), lies in the 
expected range for dwarf galaxies. Comparisons of profiles  
in the R, B, U and FUV bands indicate that profiles retain the same 
general shape while inflections due to star formation increase 
as the wavelength of the profile decreases. The radial continuity of 
star formation sites demonstrates that the time averaged level of 
activity in the past correlates with the current activity at the
same radius, supporting the theory that star formation in irregular
galaxies is dominated by local processes. 

(3) Star formation rates derived from \ha and FUV photometry are
used to characterize the current (traced by \ha) and recent (traced by FUV) 
global star formation activity. Comparisons of the two rates 
allow a rough estimate of the recent star formation history and indicate 
that the galaxy is sustaining a somewhat recent 
star formation event.

(4) Ages for individual regions are derived using B, H$\alpha$, and 
FUV photometry and show both older, diffuse FUV regions and younger, 
compact HII regions. The distribution of ages is reconciled with the 
HI morphology, showing a clear preference of young regions for
areas of dense HI and older regions for HI voids. 

(5) Local star formation triggering mechanisms are identified 
via the relative \ha and FUV morphologies. Observational evidence
of the influence that massive stars have on their surroundings
includes massive stars inside ionized shells and compact HII 
regions surrounding aging clusters. Causal relationships
between massive star formation and secondary sites of star
formation are shown.
Star formation progression
patterns are consistent with the SSPSF model in that progressively 
older chains of star formation are seen in clusters. 

(6) Large-scale star formation triggering by HI shells generated 
by massive stars is seen by comparing the FUV and HI morphologies.
Given the timescale over which FUV observations trace massive star
formation, both secondary sites of star formation and progenitor
populations are identified. Analysis of the energy available
from massive stars inside HI voids indicates that energy deposited 
into the ISM from supernovae and stellar winds are sufficient to 
account for the observed properties of the HI holes. 

(7) Global kinematical properties may also play a role in the 
star formation process since differences in the rotation characteristics
of the neutral gas disk are coupled with differences in triggering 
mechanisms. Large-scale feedback from massive star formation is shown to 
operate in regions that lack differential shear in the gas disk.

\acknowledgments
S.G.S. wishes to acknowledge the funding of this study
through 
the Alabama Space Grant Consortium project and to thank the University
of Alabama, NASA/Goddard Space Flight Center, and the 
U. S. Naval Observatory for their support. 
Funding for the UIT project has been through the Spacelab Office
at NASA Headquarters under project number 440-51.


\begin{thebibliography}{}
\bibitem[Allen \eans 1987]{1} Allen, R. J., Knapen, J. H., Bohlin, R., \&
Stecher, T. P. 1997, \apj, 487, 171

\bibitem[Blaauw 1991]{2} Blaauw, A. 1991, in {\em The Physics of Star Formation and Early Stellar Evolution}, ed. C. J. Lada \& N. D. Kylafis, (Dordrecht: Kluwer), 125 

\bibitem[Brosch, Heller, \& Almoznino 1998]{3}  Brosch, N., Heller, H., \& Almoznino, E. 1998, \mnras, 300, 1091

\bibitem[Byrd \& Howard 1992]{4}   Byrd, G. G. \& Howard, S. 1992, \aj, 103, 1089

\bibitem[Corwin 1997]{5}  Corwin, H. 1997, private communication

\bibitem[de Vaucouleurs, G. \eans 1991]{6}  de Vaucouleurs, G., de Vaucouleurs, A., 
Corwin, H. G., Buta, R. J., Paturel, G., \& Fouqu\'{e}, P. 1991, 
Third Reference Catalog of Bright Galaxies (RC3) (New York: Springer)

\bibitem[Deharveng \eans 1994]{7}  Deharveng, J. -M., Sasseen, T. P., Buat, V., 
Bowyer, S., Lampton, M., \& Wu, X. 1994, \aap, 289, 715

\bibitem[Dopita \eans 1985]{8}  Dopita, M. A., Mathewson, D. S., \& Ford, V. L. 1985, 
\apj, 297, 599

\bibitem[Dressel \& Condon 1976]{9}  Dressel, L. L. \& Condon, J. J. 1976, \apjs, 31, 187

\bibitem[Efremov \& Elmegreen 1998]{10}  Efremov, Y. N. \& Elmegreen, B. G. 1998, \mnras, 299, 643


\bibitem[Elmegreen 1992]{11}  Elmegreen, B. G. 1992, in {\em Star Formation in
Stellar Systems}, ed. G. Tenorio-Tagle \& F. S\'{a}nchez (Cambridge: Cambridge University Press), 383

\bibitem[Elmegreen \& Lada 1977]{12}  Elmegreen, B. G. \& Lada, C. J. 1977, \apj, 214, 725

\bibitem[Fanelli \eans 1997b]{13}   Fanelli, M. N., \ea 1997a, \baas, 191, \# 82.01

\bibitem[Fanelli \eans 1992]{14}   Fanelli, M. N., O'Connell, R. W., Burstein, D., \& Chi-Chao, W. 1992, \apjs, 82, 197

\bibitem[Fanelli \eans 1997a]{15}  Fanelli, M. N., \ea 1997b, in {\it Star Formation Near and Far}, ed. S. S. Holt \& L. G. Mundy (AIP Press), 598

\bibitem[Freedman 1970]{16}  Freeman, K. C. 1970, \apj, 160, 811

\bibitem[Freedman \eans 1994]{17}   Freedman, W. L., \ea 1994, \apj, 427, 628

\bibitem[Gerola \& Seiden 1978]{18}  Gerola, H. \& Seiden, P. E. 1978, \apj, 223, 129

\bibitem[Gerola \eans 1980]{19}  Gerola, H., Seiden, P. E., \& Schulman, L. S. 1980, 
\apj, 242, 517

\bibitem[Gessner \eans 1995]{20}  Gessner, S. E., \ea 1995, \baas, 187,\# 11.06

\bibitem[Heydari-Malayeri \& Testor 1981]{21}  Heydari-Malayeri, M. \& Testor, G. 
1981, \aap, 96, 219

\bibitem[Hill \eans 1995]{22}   Hill, J. K., \ea 1995, \apj, 438, 181

\bibitem[Hill \eans 1993]{23}  Hill, J. K., \ea 1993, \apjl, 402, L45

\bibitem[Hill \eans 1994]{24}   Hill, J. K., Isensee, J. E., Cornett, R. H., Bohlin, R. C., O'Connell, R. W., Roberts, M. S., Smith, A. M., \& Stecher, T. P. 1994, \apj, 425, 122

\bibitem[Hill \eans 1997]{25}   Hill, J. K., \ea 1997, \apj, 477, 673

\bibitem[Hill \eans 1998]{26}   Hill, R. S., \ea 1998, \apj, 507, 179

\bibitem[Hodge \eans 1994]{27}  Hodge, P., Strobel, N., \& Kennicutt, R. C. 
1994, \pasp, 106, 309

\bibitem[Hoessel \eans 1998]{28}  Hoessel, J. G., Saha, A., 
\& Danielson, G. E. 1998, \aj, 115, 573

\bibitem[Huchtmeier \& Richter 1988]{29}  Huchtmeier, W. K., \& Richter, O.-G.
1988, \aap, 203, 237

\bibitem[Hunter 1992]{30}  Hunter, D. A. 1992, in {\em Star Formation in
Stellar Systems}, ed. G. Tenorio-Tagle \& F. S\'{a}nchez (Cambridge: Cambridge University Press), 66

\bibitem[Hunter 1997]{31}  Hunter, D. A. 1997, \pasp, 109, 937

\bibitem[Hunter \eans 1995]{32}  Hunter, D. A., Boyd, D. M., \& Hawley, W. N.
1995, \apjs, 99, 551

\bibitem[Hunter \eans 1998]{33} Hunter, D. A., Elmegreen, B. G., \& Baker, A. L. 1998, \apj, 493, 595
 
\bibitem[Hunter \& Gallagher 1985]{34}   Hunter, D. A. \& Gallagher, J. S. 
 1985, \apjs, 58, 533

\bibitem[Hunter \eans 1989]{35}  Hunter, D. A., Gallagher, J. S., Rice, W. L., \& 
Gillett, F. C. 1989, \apj, 336, 152

\bibitem[Karachentseva \eans 1996]{36}  Karachentseva, V. E., Prugniel, P., Vennik, J., 
Richter, G. M., Thaun, T. X., \& Martin, J. M. 1996, \aaps, 117, 343

\bibitem[Kennicutt 1998]{37}  Kennicutt, R. C. 1998, \araa, 36 

\bibitem[Kim \eans 1998]{38} Kim, S., \ea 1998, \apj, 503, 674

\bibitem[Kurucz 1992]{39}  Kurucz, R. L. 1992, in {\em Stellar Populations of Galaxies},
eds. B. Barbuy \& A. Renzini, (Dordrecht:Kluwer), 225

\bibitem[Lafon \eans 1983]{40}  Lafon, G., Deharveng, L., Baudry, A., \&
de La N\"{o}e, J. 1983, \aap, 124, 1

\bibitem[Leitherer \& Heckman 1995]{41}  Leitherer, C. \& Heckman, T. M. 1995, \apjs, 96, 9

\bibitem[Lortet \& Testor 1988]{42}  Lortet, M.-C. \& Testor, G. 1988, \aap, 194, 11

\bibitem[Madau \eans 1998]{43}  Madau, P., Pozzetti, L., \& Dickinson, M. 1998, \apj, 498, 106

\bibitem[Maschenko \& Silich 1995]{44}  Maschenko, S. Y. \& Silich, S. A. 1995, 
Astronomy Reports, 39, 587

\bibitem[Massey \eans 1996]{45}  Massey, P., Bianchi, L., Hutchings, J. B., \& Stecher, T. P. 1996, \apj, 469, 629 

\bibitem[Melisse \& Isreal 1994]{46} Melisse, J. P. M. \& Israel, F. P. 1994, \aap, 285, 51

\bibitem[Miller \& Hodge 1994]{47}  Miller, B. W. \& Hodge, P. 1994, \apj, 427, 656

\bibitem[Nilson 1973]{48}  Nilson, P. 1973, Uppsala General Catalogue of Galaxies 
(Uppsala Astr. Obs. Ann., 6)

\bibitem[O'Connell 1997]{49}   O'Connell, R. W. 1997, in 
{\it The Ultraviolet Universe at Low and High Redshift}, ed. W. H. Waller, Fanelli, M. N., 
Hollis, J. E., \& Danks, A. C., (AIP Press), 11

\bibitem[Page \& Carruthers 1981]{50}  Page, T. \& Carruthers, G. R. 1981, \apj, 248, 906

\bibitem[Parker 1988]{51}  Parker, J. Wm. 1998, \aj, 116, 180

\bibitem[Patterson \& Thuan 1996]{52}  Patterson, R. J. \& Thuan, T. X. 1996, \aaps,
107, 103

\bibitem[Puche \eans 1992]{53}  Puche, D., Westpfahl, D., Brinks, E., \& Roy, J.-R.
1992, \aj, 103, 1841 

\bibitem[Rhode \eans 1998]{54}  Rhode, K. L., Salzer, J. J.,
 Westpfahl, D. J., \& Radice, L. A. 1999, \aj, 118, 323

\bibitem[R\"{o}nnback \& Bergvall 1994]{55}  R\"{o}nnback, J. and Bergvall,
N. 1994, \aaps, 108, 193

\bibitem[Ryder \& Dopita 1994]{56}  Ryder, S. D. \& Dopita, M. A. 1994, \apj, 430, 142

\bibitem[Salpeter 1955]{57}  Salpeter, E. E. 1955, \apj, 121, 161

\bibitem[Savage \& Mathis 1979]{58}  Savage, B. D. \& Mathis, J. S. \araa, 17, 73

\bibitem[Schaerer 1993]{59}  Schaerer, D., Meynet, G., Maeder, A., Schaller, G.
1993, \aaps, 98, 523

\bibitem[Stecher \eans 1992]{60}  Stecher, T. P., \ea 1992, \apjl, 395, L1

\bibitem[Stecher \eans 1997]{61}  Stecher, T. P., \ea 1997, \pasp, 109, 735

\bibitem[Stewart 1998]{62}  Stewart, S. G. 1998, Ph.D. Thesis, University of Alabama

\bibitem[Stone 1977]{63}  Stone, R. P. S. 1977, \apj, 218, 767

\bibitem[Tacconi \& Young 1987]{64}  Tacconi, L. J. \& Young, J. S. 1987, \apj, 322, 681

\bibitem[Tenorio-Tagle 1979]{65}  Tenorio-Tagle, G. 1979, \aap, 71, 59

\bibitem[Tenorio-Tagle \& Bodenheimer 1988]{66}  Tenorio-Tagle, G.
 \& Bodenheimer, P. 1988, \araa, 26, 145

\bibitem[Tongue \& Westpfahl 1995]{67}  Tongue, T. D. \& Westpfahl, D. J.
1995, \aj, 109, 2462

\bibitem[Vacca \eans 1995]{68}  Vacca, W. D., Robert, C., Leitherer, C., \& Conti, P. S. 1995, \apj, 444, 647

\bibitem[Vader \& Chaboyer 1994]{69}  Vader, J. P. \& Chaboyer, B. 1994, \aj, 108, 1209

\bibitem[van den Bergh 1959]{70}  van den Bergh, S.  1959, Publ. David Dunlap Obs. II, No. 5

\bibitem[van der Hulst \& Sancisi 1988]{71}  van der Hulst, T. \& Sancisi, R. 1988,
\aj, 95, 1354

\bibitem[Van Dyk, Puche, \& Wong 1998]{72} Van Dyk, S. D., Puche, D., \& Wong, T. 1998, \aj, 116, 2341

\bibitem[Verter \& Rickard 1998]{73} Verter, F. \& Rickard, L. J. 1998, \aj, 115, 745

\bibitem[Waller 1990]{74}  Waller, W. 1990, \pasp, 102, 1217

\bibitem[Weaver \eans 1977]{75}  Weaver, R., McCray, R., Castor, J., Shapiro, P., \&
Moore, R. 1977, \apj, 218, 377

\bibitem[Westpfahl 1997]{76}  Westpfahl, D. 1997, Private Communication

\end{thebibliography}
\end{document}